\documentclass[a4paper,amsmath,amssymb]{iopart}

\usepackage{color}

\definecolor{AliceBlue}{rgb}{0.94,0.97,1.00}
\definecolor{AntiqueWhite1}{rgb}{1.00,0.94,0.86}
\definecolor{AntiqueWhite2}{rgb}{0.93,0.87,0.80}
\definecolor{AntiqueWhite3}{rgb}{0.80,0.75,0.69}
\definecolor{AntiqueWhite4}{rgb}{0.55,0.51,0.47}
\definecolor{AntiqueWhite}{rgb}{0.98,0.92,0.84}
\definecolor{BlanchedAlmond}{rgb}{1.00,0.92,0.80}
\definecolor{BlueViolet}{rgb}{0.54,0.17,0.89}
\definecolor{CadetBlue1}{rgb}{0.60,0.96,1.00}
\definecolor{CadetBlue2}{rgb}{0.56,0.90,0.93}
\definecolor{CadetBlue3}{rgb}{0.48,0.77,0.80}
\definecolor{CadetBlue4}{rgb}{0.33,0.53,0.55}
\definecolor{CadetBlue}{rgb}{0.37,0.62,0.63}
\definecolor{CornflowerBlue}{rgb}{0.39,0.58,0.93}
\definecolor{DarkBlue}{rgb}{0.00,0.00,0.55}
\definecolor{DarkCyan}{rgb}{0.00,0.55,0.55}
\definecolor{DarkGoldenrod1}{rgb}{1.00,0.73,0.06}
\definecolor{DarkGoldenrod2}{rgb}{0.93,0.68,0.05}
\definecolor{DarkGoldenrod3}{rgb}{0.80,0.58,0.05}
\definecolor{DarkGoldenrod4}{rgb}{0.55,0.40,0.03}
\definecolor{DarkGoldenrod}{rgb}{0.72,0.53,0.04}
\definecolor{DarkGray}{rgb}{0.66,0.66,0.66}
\definecolor{DarkGreen}{rgb}{0.00,0.39,0.00}
\definecolor{DarkGrey}{rgb}{0.66,0.66,0.66}
\definecolor{DarkKhaki}{rgb}{0.74,0.72,0.42}
\definecolor{DarkMagenta}{rgb}{0.55,0.00,0.55}
\definecolor{DarkOliveGreen1}{rgb}{0.79,1.00,0.44}
\definecolor{DarkOliveGreen2}{rgb}{0.74,0.93,0.41}
\definecolor{DarkOliveGreen3}{rgb}{0.64,0.80,0.35}
\definecolor{DarkOliveGreen4}{rgb}{0.43,0.55,0.24}
\definecolor{DarkOliveGreen}{rgb}{0.33,0.42,0.18}
\definecolor{DarkOrange1}{rgb}{1.00,0.50,0.00}
\definecolor{DarkOrange2}{rgb}{0.93,0.46,0.00}
\definecolor{DarkOrange3}{rgb}{0.80,0.40,0.00}
\definecolor{DarkOrange4}{rgb}{0.55,0.27,0.00}
\definecolor{DarkOrange}{rgb}{1.00,0.55,0.00}
\definecolor{DarkOrchid1}{rgb}{0.75,0.24,1.00}
\definecolor{DarkOrchid2}{rgb}{0.70,0.23,0.93}
\definecolor{DarkOrchid3}{rgb}{0.60,0.20,0.80}
\definecolor{DarkOrchid4}{rgb}{0.41,0.13,0.55}
\definecolor{DarkOrchid}{rgb}{0.60,0.20,0.80}
\definecolor{DarkRed}{rgb}{0.55,0.00,0.00}
\definecolor{DarkSalmon}{rgb}{0.91,0.59,0.48}
\definecolor{DarkSeaGreen1}{rgb}{0.76,1.00,0.76}
\definecolor{DarkSeaGreen2}{rgb}{0.71,0.93,0.71}
\definecolor{DarkSeaGreen3}{rgb}{0.61,0.80,0.61}
\definecolor{DarkSeaGreen4}{rgb}{0.41,0.55,0.41}
\definecolor{DarkSeaGreen}{rgb}{0.56,0.74,0.56}
\definecolor{DarkSlateBlue}{rgb}{0.28,0.24,0.55}
\definecolor{DarkSlateGray1}{rgb}{0.59,1.00,1.00}
\definecolor{DarkSlateGray2}{rgb}{0.55,0.93,0.93}
\definecolor{DarkSlateGray3}{rgb}{0.47,0.80,0.80}
\definecolor{DarkSlateGray4}{rgb}{0.32,0.55,0.55}
\definecolor{DarkSlateGray}{rgb}{0.18,0.31,0.31}
\definecolor{DarkSlateGrey}{rgb}{0.18,0.31,0.31}
\definecolor{DarkTurquoise}{rgb}{0.00,0.81,0.82}
\definecolor{DarkViolet}{rgb}{0.58,0.00,0.83}
\definecolor{DeepPink1}{rgb}{1.00,0.08,0.58}
\definecolor{DeepPink2}{rgb}{0.93,0.07,0.54}
\definecolor{DeepPink3}{rgb}{0.80,0.06,0.46}
\definecolor{DeepPink4}{rgb}{0.55,0.04,0.31}
\definecolor{DeepPink}{rgb}{1.00,0.08,0.58}
\definecolor{DeepSkyBlue1}{rgb}{0.00,0.75,1.00}
\definecolor{DeepSkyBlue2}{rgb}{0.00,0.70,0.93}
\definecolor{DeepSkyBlue3}{rgb}{0.00,0.60,0.80}
\definecolor{DeepSkyBlue4}{rgb}{0.00,0.41,0.55}
\definecolor{DeepSkyBlue}{rgb}{0.00,0.75,1.00}
\definecolor{DimGray}{rgb}{0.41,0.41,0.41}
\definecolor{DimGrey}{rgb}{0.41,0.41,0.41}
\definecolor{DodgerBlue1}{rgb}{0.12,0.56,1.00}
\definecolor{DodgerBlue2}{rgb}{0.11,0.53,0.93}
\definecolor{DodgerBlue3}{rgb}{0.09,0.45,0.80}
\definecolor{DodgerBlue4}{rgb}{0.06,0.31,0.55}
\definecolor{DodgerBlue}{rgb}{0.12,0.56,1.00}
\definecolor{FloralWhite}{rgb}{1.00,0.98,0.94}
\definecolor{ForestGreen}{rgb}{0.13,0.55,0.13}
\definecolor{GhostWhite}{rgb}{0.97,0.97,1.00}
\definecolor{GreenYellow}{rgb}{0.68,1.00,0.18}
\definecolor{HotPink1}{rgb}{1.00,0.43,0.71}
\definecolor{HotPink2}{rgb}{0.93,0.42,0.65}
\definecolor{HotPink3}{rgb}{0.80,0.38,0.56}
\definecolor{HotPink4}{rgb}{0.55,0.23,0.38}
\definecolor{HotPink}{rgb}{1.00,0.41,0.71}
\definecolor{IndianRed1}{rgb}{1.00,0.42,0.42}
\definecolor{IndianRed2}{rgb}{0.93,0.39,0.39}
\definecolor{IndianRed3}{rgb}{0.80,0.33,0.33}
\definecolor{IndianRed4}{rgb}{0.55,0.23,0.23}
\definecolor{IndianRed}{rgb}{0.80,0.36,0.36}
\definecolor{LavenderBlush1}{rgb}{1.00,0.94,0.96}
\definecolor{LavenderBlush2}{rgb}{0.93,0.88,0.90}
\definecolor{LavenderBlush3}{rgb}{0.80,0.76,0.77}
\definecolor{LavenderBlush4}{rgb}{0.55,0.51,0.53}
\definecolor{LavenderBlush}{rgb}{1.00,0.94,0.96}
\definecolor{LawnGreen}{rgb}{0.49,0.99,0.00}
\definecolor{LemonChiffon1}{rgb}{1.00,0.98,0.80}
\definecolor{LemonChiffon2}{rgb}{0.93,0.91,0.75}
\definecolor{LemonChiffon3}{rgb}{0.80,0.79,0.65}
\definecolor{LemonChiffon4}{rgb}{0.55,0.54,0.44}
\definecolor{LemonChiffon}{rgb}{1.00,0.98,0.80}
\definecolor{LightBlue1}{rgb}{0.75,0.94,1.00}
\definecolor{LightBlue2}{rgb}{0.70,0.87,0.93}
\definecolor{LightBlue3}{rgb}{0.60,0.75,0.80}
\definecolor{LightBlue4}{rgb}{0.41,0.51,0.55}
\definecolor{LightBlue}{rgb}{0.68,0.85,0.90}
\definecolor{LightCoral}{rgb}{0.94,0.50,0.50}
\definecolor{LightCyan1}{rgb}{0.88,1.00,1.00}
\definecolor{LightCyan2}{rgb}{0.82,0.93,0.93}
\definecolor{LightCyan3}{rgb}{0.71,0.80,0.80}
\definecolor{LightCyan4}{rgb}{0.48,0.55,0.55}
\definecolor{LightCyan}{rgb}{0.88,1.00,1.00}
\definecolor{LightGoldenrod1}{rgb}{1.00,0.93,0.55}
\definecolor{LightGoldenrod2}{rgb}{0.93,0.86,0.51}
\definecolor{LightGoldenrod3}{rgb}{0.80,0.75,0.44}
\definecolor{LightGoldenrod4}{rgb}{0.55,0.51,0.30}
\definecolor{LightGoldenrodYellow}{rgb}{0.98,0.98,0.82}
\definecolor{LightGoldenrod}{rgb}{0.93,0.87,0.51}
\definecolor{LightGray}{rgb}{0.83,0.83,0.83}
\definecolor{LightGreen}{rgb}{0.56,0.93,0.56}
\definecolor{LightGrey}{rgb}{0.83,0.83,0.83}
\definecolor{LightPink1}{rgb}{1.00,0.68,0.73}
\definecolor{LightPink2}{rgb}{0.93,0.64,0.68}
\definecolor{LightPink3}{rgb}{0.80,0.55,0.58}
\definecolor{LightPink4}{rgb}{0.55,0.37,0.40}
\definecolor{LightPink}{rgb}{1.00,0.71,0.76}
\definecolor{LightSalmon1}{rgb}{1.00,0.63,0.48}
\definecolor{LightSalmon2}{rgb}{0.93,0.58,0.45}
\definecolor{LightSalmon3}{rgb}{0.80,0.51,0.38}
\definecolor{LightSalmon4}{rgb}{0.55,0.34,0.26}
\definecolor{LightSalmon}{rgb}{1.00,0.63,0.48}
\definecolor{LightSeaGreen}{rgb}{0.13,0.70,0.67}
\definecolor{LightSkyBlue1}{rgb}{0.69,0.89,1.00}
\definecolor{LightSkyBlue2}{rgb}{0.64,0.83,0.93}
\definecolor{LightSkyBlue3}{rgb}{0.55,0.71,0.80}
\definecolor{LightSkyBlue4}{rgb}{0.38,0.48,0.55}
\definecolor{LightSkyBlue}{rgb}{0.53,0.81,0.98}
\definecolor{LightSlateBlue}{rgb}{0.52,0.44,1.00}
\definecolor{LightSlateGray}{rgb}{0.47,0.53,0.60}
\definecolor{LightSlateGrey}{rgb}{0.47,0.53,0.60}
\definecolor{LightSteelBlue1}{rgb}{0.79,0.88,1.00}
\definecolor{LightSteelBlue2}{rgb}{0.74,0.82,0.93}
\definecolor{LightSteelBlue3}{rgb}{0.64,0.71,0.80}
\definecolor{LightSteelBlue4}{rgb}{0.43,0.48,0.55}
\definecolor{LightSteelBlue}{rgb}{0.69,0.77,0.87}
\definecolor{LightYellow1}{rgb}{1.00,1.00,0.88}
\definecolor{LightYellow2}{rgb}{0.93,0.93,0.82}
\definecolor{LightYellow3}{rgb}{0.80,0.80,0.71}
\definecolor{LightYellow4}{rgb}{0.55,0.55,0.48}
\definecolor{LightYellow}{rgb}{1.00,1.00,0.88}
\definecolor{LimeGreen}{rgb}{0.20,0.80,0.20}
\definecolor{MediumAquamarine}{rgb}{0.40,0.80,0.67}
\definecolor{MediumBlue}{rgb}{0.00,0.00,0.80}
\definecolor{MediumOrchid1}{rgb}{0.88,0.40,1.00}
\definecolor{MediumOrchid2}{rgb}{0.82,0.37,0.93}
\definecolor{MediumOrchid3}{rgb}{0.71,0.32,0.80}
\definecolor{MediumOrchid4}{rgb}{0.48,0.22,0.55}
\definecolor{MediumOrchid}{rgb}{0.73,0.33,0.83}
\definecolor{MediumPurple1}{rgb}{0.67,0.51,1.00}
\definecolor{MediumPurple2}{rgb}{0.62,0.47,0.93}
\definecolor{MediumPurple3}{rgb}{0.54,0.41,0.80}
\definecolor{MediumPurple4}{rgb}{0.36,0.28,0.55}
\definecolor{MediumPurple}{rgb}{0.58,0.44,0.86}
\definecolor{MediumSeaGreen}{rgb}{0.24,0.70,0.44}
\definecolor{MediumSlateBlue}{rgb}{0.48,0.41,0.93}
\definecolor{MediumSpringGreen}{rgb}{0.00,0.98,0.60}
\definecolor{MediumTurquoise}{rgb}{0.28,0.82,0.80}
\definecolor{MediumVioletRed}{rgb}{0.78,0.08,0.52}
\definecolor{MidnightBlue}{rgb}{0.10,0.10,0.44}
\definecolor{MintCream}{rgb}{0.96,1.00,0.98}
\definecolor{MistyRose1}{rgb}{1.00,0.89,0.88}
\definecolor{MistyRose2}{rgb}{0.93,0.84,0.82}
\definecolor{MistyRose3}{rgb}{0.80,0.72,0.71}
\definecolor{MistyRose4}{rgb}{0.55,0.49,0.48}
\definecolor{MistyRose}{rgb}{1.00,0.89,0.88}
\definecolor{NavajoWhite1}{rgb}{1.00,0.87,0.68}
\definecolor{NavajoWhite2}{rgb}{0.93,0.81,0.63}
\definecolor{NavajoWhite3}{rgb}{0.80,0.70,0.55}
\definecolor{NavajoWhite4}{rgb}{0.55,0.47,0.37}
\definecolor{NavajoWhite}{rgb}{1.00,0.87,0.68}
\definecolor{NavyBlue}{rgb}{0.00,0.00,0.50}
\definecolor{OldLace}{rgb}{0.99,0.96,0.90}
\definecolor{OliveDrab1}{rgb}{0.75,1.00,0.24}
\definecolor{OliveDrab2}{rgb}{0.70,0.93,0.23}
\definecolor{OliveDrab3}{rgb}{0.60,0.80,0.20}
\definecolor{OliveDrab4}{rgb}{0.41,0.55,0.13}
\definecolor{OliveDrab}{rgb}{0.42,0.56,0.14}
\definecolor{OrangeRed1}{rgb}{1.00,0.27,0.00}
\definecolor{OrangeRed2}{rgb}{0.93,0.25,0.00}
\definecolor{OrangeRed3}{rgb}{0.80,0.22,0.00}
\definecolor{OrangeRed4}{rgb}{0.55,0.15,0.00}
\definecolor{OrangeRed}{rgb}{1.00,0.27,0.00}
\definecolor{PaleGoldenrod}{rgb}{0.93,0.91,0.67}
\definecolor{PaleGreen1}{rgb}{0.60,1.00,0.60}
\definecolor{PaleGreen2}{rgb}{0.56,0.93,0.56}
\definecolor{PaleGreen3}{rgb}{0.49,0.80,0.49}
\definecolor{PaleGreen4}{rgb}{0.33,0.55,0.33}
\definecolor{PaleGreen}{rgb}{0.60,0.98,0.60}
\definecolor{PaleTurquoise1}{rgb}{0.73,1.00,1.00}
\definecolor{PaleTurquoise2}{rgb}{0.68,0.93,0.93}
\definecolor{PaleTurquoise3}{rgb}{0.59,0.80,0.80}
\definecolor{PaleTurquoise4}{rgb}{0.40,0.55,0.55}
\definecolor{PaleTurquoise}{rgb}{0.69,0.93,0.93}
\definecolor{PaleVioletRed1}{rgb}{1.00,0.51,0.67}
\definecolor{PaleVioletRed2}{rgb}{0.93,0.47,0.62}
\definecolor{PaleVioletRed3}{rgb}{0.80,0.41,0.54}
\definecolor{PaleVioletRed4}{rgb}{0.55,0.28,0.36}
\definecolor{PaleVioletRed}{rgb}{0.86,0.44,0.58}
\definecolor{PapayaWhip}{rgb}{1.00,0.94,0.84}
\definecolor{PeachPuff1}{rgb}{1.00,0.85,0.73}
\definecolor{PeachPuff2}{rgb}{0.93,0.80,0.68}
\definecolor{PeachPuff3}{rgb}{0.80,0.69,0.58}
\definecolor{PeachPuff4}{rgb}{0.55,0.47,0.40}
\definecolor{PeachPuff}{rgb}{1.00,0.85,0.73}
\definecolor{PowderBlue}{rgb}{0.69,0.88,0.90}
\definecolor{RosyBrown1}{rgb}{1.00,0.76,0.76}
\definecolor{RosyBrown2}{rgb}{0.93,0.71,0.71}
\definecolor{RosyBrown3}{rgb}{0.80,0.61,0.61}
\definecolor{RosyBrown4}{rgb}{0.55,0.41,0.41}
\definecolor{RosyBrown}{rgb}{0.74,0.56,0.56}
\definecolor{RoyalBlue1}{rgb}{0.28,0.46,1.00}
\definecolor{RoyalBlue2}{rgb}{0.26,0.43,0.93}
\definecolor{RoyalBlue3}{rgb}{0.23,0.37,0.80}
\definecolor{RoyalBlue4}{rgb}{0.15,0.25,0.55}
\definecolor{RoyalBlue}{rgb}{0.25,0.41,0.88}
\definecolor{SaddleBrown}{rgb}{0.55,0.27,0.07}
\definecolor{SandyBrown}{rgb}{0.96,0.64,0.38}
\definecolor{SeaGreen1}{rgb}{0.33,1.00,0.62}
\definecolor{SeaGreen2}{rgb}{0.31,0.93,0.58}
\definecolor{SeaGreen3}{rgb}{0.26,0.80,0.50}
\definecolor{SeaGreen4}{rgb}{0.18,0.55,0.34}
\definecolor{SeaGreen}{rgb}{0.18,0.55,0.34}
\definecolor{SkyBlue1}{rgb}{0.53,0.81,1.00}
\definecolor{SkyBlue2}{rgb}{0.49,0.75,0.93}
\definecolor{SkyBlue3}{rgb}{0.42,0.65,0.80}
\definecolor{SkyBlue4}{rgb}{0.29,0.44,0.55}
\definecolor{SkyBlue}{rgb}{0.53,0.81,0.92}
\definecolor{SlateBlue1}{rgb}{0.51,0.44,1.00}
\definecolor{SlateBlue2}{rgb}{0.48,0.40,0.93}
\definecolor{SlateBlue3}{rgb}{0.41,0.35,0.80}
\definecolor{SlateBlue4}{rgb}{0.28,0.24,0.55}
\definecolor{SlateBlue}{rgb}{0.42,0.35,0.80}
\definecolor{SlateGray1}{rgb}{0.78,0.89,1.00}
\definecolor{SlateGray2}{rgb}{0.73,0.83,0.93}
\definecolor{SlateGray3}{rgb}{0.62,0.71,0.80}
\definecolor{SlateGray4}{rgb}{0.42,0.48,0.55}
\definecolor{SlateGray}{rgb}{0.44,0.50,0.56}
\definecolor{SlateGrey}{rgb}{0.44,0.50,0.56}
\definecolor{SpringGreen1}{rgb}{0.00,1.00,0.50}
\definecolor{SpringGreen2}{rgb}{0.00,0.93,0.46}
\definecolor{SpringGreen3}{rgb}{0.00,0.80,0.40}
\definecolor{SpringGreen4}{rgb}{0.00,0.55,0.27}
\definecolor{SpringGreen}{rgb}{0.00,1.00,0.50}
\definecolor{SteelBlue1}{rgb}{0.39,0.72,1.00}
\definecolor{SteelBlue2}{rgb}{0.36,0.67,0.93}
\definecolor{SteelBlue3}{rgb}{0.31,0.58,0.80}
\definecolor{SteelBlue4}{rgb}{0.21,0.39,0.55}
\definecolor{SteelBlue}{rgb}{0.27,0.51,0.71}
\definecolor{VioletRed1}{rgb}{1.00,0.24,0.59}
\definecolor{VioletRed2}{rgb}{0.93,0.23,0.55}
\definecolor{VioletRed3}{rgb}{0.80,0.20,0.47}
\definecolor{VioletRed4}{rgb}{0.55,0.13,0.32}
\definecolor{VioletRed}{rgb}{0.82,0.13,0.56}
\definecolor{WhiteSmoke}{rgb}{0.96,0.96,0.96}
\definecolor{YellowGreen}{rgb}{0.60,0.80,0.20}
\definecolor{aliceblue}{rgb}{0.94,0.97,1.00}
\definecolor{antiquewhite}{rgb}{0.98,0.92,0.84}
\definecolor{aquamarine1}{rgb}{0.50,1.00,0.83}
\definecolor{aquamarine2}{rgb}{0.46,0.93,0.78}
\definecolor{aquamarine3}{rgb}{0.40,0.80,0.67}
\definecolor{aquamarine4}{rgb}{0.27,0.55,0.45}
\definecolor{aquamarine}{rgb}{0.50,1.00,0.83}
\definecolor{azure1}{rgb}{0.94,1.00,1.00}
\definecolor{azure2}{rgb}{0.88,0.93,0.93}
\definecolor{azure3}{rgb}{0.76,0.80,0.80}
\definecolor{azure4}{rgb}{0.51,0.55,0.55}
\definecolor{azure}{rgb}{0.94,1.00,1.00}
\definecolor{beige}{rgb}{0.96,0.96,0.86}
\definecolor{bisque1}{rgb}{1.00,0.89,0.77}
\definecolor{bisque2}{rgb}{0.93,0.84,0.72}
\definecolor{bisque3}{rgb}{0.80,0.72,0.62}
\definecolor{bisque4}{rgb}{0.55,0.49,0.42}
\definecolor{bisque}{rgb}{1.00,0.89,0.77}
\definecolor{black}{rgb}{0.00,0.00,0.00}
\definecolor{blanchedalmond}{rgb}{1.00,0.92,0.80}
\definecolor{blue1}{rgb}{0.00,0.00,1.00}
\definecolor{blue2}{rgb}{0.00,0.00,0.93}
\definecolor{blue3}{rgb}{0.00,0.00,0.80}
\definecolor{blue4}{rgb}{0.00,0.00,0.55}
\definecolor{blueviolet}{rgb}{0.54,0.17,0.89}
\definecolor{blue}{rgb}{0.00,0.00,1.00}
\definecolor{brown1}{rgb}{1.00,0.25,0.25}
\definecolor{brown2}{rgb}{0.93,0.23,0.23}
\definecolor{brown3}{rgb}{0.80,0.20,0.20}
\definecolor{brown4}{rgb}{0.55,0.14,0.14}
\definecolor{brown}{rgb}{0.65,0.16,0.16}
\definecolor{burlywood1}{rgb}{1.00,0.83,0.61}
\definecolor{burlywood2}{rgb}{0.93,0.77,0.57}
\definecolor{burlywood3}{rgb}{0.80,0.67,0.49}
\definecolor{burlywood4}{rgb}{0.55,0.45,0.33}
\definecolor{burlywood}{rgb}{0.87,0.72,0.53}
\definecolor{cadetblue}{rgb}{0.37,0.62,0.63}
\definecolor{chartreuse1}{rgb}{0.50,1.00,0.00}
\definecolor{chartreuse2}{rgb}{0.46,0.93,0.00}
\definecolor{chartreuse3}{rgb}{0.40,0.80,0.00}
\definecolor{chartreuse4}{rgb}{0.27,0.55,0.00}
\definecolor{chartreuse}{rgb}{0.50,1.00,0.00}
\definecolor{chocolate1}{rgb}{1.00,0.50,0.14}
\definecolor{chocolate2}{rgb}{0.93,0.46,0.13}
\definecolor{chocolate3}{rgb}{0.80,0.40,0.11}
\definecolor{chocolate4}{rgb}{0.55,0.27,0.07}
\definecolor{chocolate}{rgb}{0.82,0.41,0.12}
\definecolor{coral1}{rgb}{1.00,0.45,0.34}
\definecolor{coral2}{rgb}{0.93,0.42,0.31}
\definecolor{coral3}{rgb}{0.80,0.36,0.27}
\definecolor{coral4}{rgb}{0.55,0.24,0.18}
\definecolor{coral}{rgb}{1.00,0.50,0.31}
\definecolor{cornflowerblue}{rgb}{0.39,0.58,0.93}
\definecolor{cornsilk1}{rgb}{1.00,0.97,0.86}
\definecolor{cornsilk2}{rgb}{0.93,0.91,0.80}
\definecolor{cornsilk3}{rgb}{0.80,0.78,0.69}
\definecolor{cornsilk4}{rgb}{0.55,0.53,0.47}
\definecolor{cornsilk}{rgb}{1.00,0.97,0.86}
\definecolor{cyan1}{rgb}{0.00,1.00,1.00}
\definecolor{cyan2}{rgb}{0.00,0.93,0.93}
\definecolor{cyan3}{rgb}{0.00,0.80,0.80}
\definecolor{cyan4}{rgb}{0.00,0.55,0.55}
\definecolor{cyan}{rgb}{0.00,1.00,1.00}
\definecolor{darkblue}{rgb}{0.00,0.00,0.55}
\definecolor{darkcyan}{rgb}{0.00,0.55,0.55}
\definecolor{darkgoldenrod}{rgb}{0.72,0.53,0.04}
\definecolor{darkgray}{rgb}{0.66,0.66,0.66}
\definecolor{darkgreen}{rgb}{0.00,0.39,0.00}
\definecolor{darkgrey}{rgb}{0.66,0.66,0.66}
\definecolor{darkkhaki}{rgb}{0.74,0.72,0.42}
\definecolor{darkmagenta}{rgb}{0.55,0.00,0.55}
\definecolor{darkolive}{rgb}{0.33,0.42,0.18}
\definecolor{darkorange}{rgb}{1.00,0.55,0.00}
\definecolor{darkorchid}{rgb}{0.60,0.20,0.80}
\definecolor{darkred}{rgb}{0.55,0.00,0.00}
\definecolor{darksalmon}{rgb}{0.91,0.59,0.48}
\definecolor{darksea}{rgb}{0.56,0.74,0.56}
\definecolor{darkslate}{rgb}{0.18,0.31,0.31}
\definecolor{darkslate}{rgb}{0.18,0.31,0.31}
\definecolor{darkslate}{rgb}{0.28,0.24,0.55}
\definecolor{darkturquoise}{rgb}{0.00,0.81,0.82}
\definecolor{darkviolet}{rgb}{0.58,0.00,0.83}
\definecolor{deeppink}{rgb}{1.00,0.08,0.58}
\definecolor{deepsky}{rgb}{0.00,0.75,1.00}
\definecolor{dimgray}{rgb}{0.41,0.41,0.41}
\definecolor{dimgrey}{rgb}{0.41,0.41,0.41}
\definecolor{dodgerblue}{rgb}{0.12,0.56,1.00}
\definecolor{firebrick1}{rgb}{1.00,0.19,0.19}
\definecolor{firebrick2}{rgb}{0.93,0.17,0.17}
\definecolor{firebrick3}{rgb}{0.80,0.15,0.15}
\definecolor{firebrick4}{rgb}{0.55,0.10,0.10}
\definecolor{firebrick}{rgb}{0.70,0.13,0.13}
\definecolor{floralwhite}{rgb}{1.00,0.98,0.94}
\definecolor{forestgreen}{rgb}{0.13,0.55,0.13}
\definecolor{gainsboro}{rgb}{0.86,0.86,0.86}
\definecolor{ghostwhite}{rgb}{0.97,0.97,1.00}
\definecolor{gold1}{rgb}{1.00,0.84,0.00}
\definecolor{gold2}{rgb}{0.93,0.79,0.00}
\definecolor{gold3}{rgb}{0.80,0.68,0.00}
\definecolor{gold4}{rgb}{0.55,0.46,0.00}
\definecolor{goldenrod1}{rgb}{1.00,0.76,0.15}
\definecolor{goldenrod2}{rgb}{0.93,0.71,0.13}
\definecolor{goldenrod3}{rgb}{0.80,0.61,0.11}
\definecolor{goldenrod4}{rgb}{0.55,0.41,0.08}
\definecolor{goldenrod}{rgb}{0.85,0.65,0.13}
\definecolor{gold}{rgb}{1.00,0.84,0.00}
\definecolor{gray0}{rgb}{0.00,0.00,0.00}
\definecolor{gray100}{rgb}{1.00,1.00,1.00}
\definecolor{gray10}{rgb}{0.10,0.10,0.10}
\definecolor{gray11}{rgb}{0.11,0.11,0.11}
\definecolor{gray12}{rgb}{0.12,0.12,0.12}
\definecolor{gray13}{rgb}{0.13,0.13,0.13}
\definecolor{gray14}{rgb}{0.14,0.14,0.14}
\definecolor{gray15}{rgb}{0.15,0.15,0.15}
\definecolor{gray16}{rgb}{0.16,0.16,0.16}
\definecolor{gray17}{rgb}{0.17,0.17,0.17}
\definecolor{gray18}{rgb}{0.18,0.18,0.18}
\definecolor{gray19}{rgb}{0.19,0.19,0.19}
\definecolor{gray1}{rgb}{0.01,0.01,0.01}
\definecolor{gray20}{rgb}{0.20,0.20,0.20}
\definecolor{gray21}{rgb}{0.21,0.21,0.21}
\definecolor{gray22}{rgb}{0.22,0.22,0.22}
\definecolor{gray23}{rgb}{0.23,0.23,0.23}
\definecolor{gray24}{rgb}{0.24,0.24,0.24}
\definecolor{gray25}{rgb}{0.25,0.25,0.25}
\definecolor{gray26}{rgb}{0.26,0.26,0.26}
\definecolor{gray27}{rgb}{0.27,0.27,0.27}
\definecolor{gray28}{rgb}{0.28,0.28,0.28}
\definecolor{gray29}{rgb}{0.29,0.29,0.29}
\definecolor{gray2}{rgb}{0.02,0.02,0.02}
\definecolor{gray30}{rgb}{0.30,0.30,0.30}
\definecolor{gray31}{rgb}{0.31,0.31,0.31}
\definecolor{gray32}{rgb}{0.32,0.32,0.32}
\definecolor{gray33}{rgb}{0.33,0.33,0.33}
\definecolor{gray34}{rgb}{0.34,0.34,0.34}
\definecolor{gray35}{rgb}{0.35,0.35,0.35}
\definecolor{gray36}{rgb}{0.36,0.36,0.36}
\definecolor{gray37}{rgb}{0.37,0.37,0.37}
\definecolor{gray38}{rgb}{0.38,0.38,0.38}
\definecolor{gray39}{rgb}{0.39,0.39,0.39}
\definecolor{gray3}{rgb}{0.03,0.03,0.03}
\definecolor{gray40}{rgb}{0.40,0.40,0.40}
\definecolor{gray41}{rgb}{0.41,0.41,0.41}
\definecolor{gray42}{rgb}{0.42,0.42,0.42}
\definecolor{gray43}{rgb}{0.43,0.43,0.43}
\definecolor{gray44}{rgb}{0.44,0.44,0.44}
\definecolor{gray45}{rgb}{0.45,0.45,0.45}
\definecolor{gray46}{rgb}{0.46,0.46,0.46}
\definecolor{gray47}{rgb}{0.47,0.47,0.47}
\definecolor{gray48}{rgb}{0.48,0.48,0.48}
\definecolor{gray49}{rgb}{0.49,0.49,0.49}
\definecolor{gray4}{rgb}{0.04,0.04,0.04}
\definecolor{gray50}{rgb}{0.50,0.50,0.50}
\definecolor{gray51}{rgb}{0.51,0.51,0.51}
\definecolor{gray52}{rgb}{0.52,0.52,0.52}
\definecolor{gray53}{rgb}{0.53,0.53,0.53}
\definecolor{gray54}{rgb}{0.54,0.54,0.54}
\definecolor{gray55}{rgb}{0.55,0.55,0.55}
\definecolor{gray56}{rgb}{0.56,0.56,0.56}
\definecolor{gray57}{rgb}{0.57,0.57,0.57}
\definecolor{gray58}{rgb}{0.58,0.58,0.58}
\definecolor{gray59}{rgb}{0.59,0.59,0.59}
\definecolor{gray5}{rgb}{0.05,0.05,0.05}
\definecolor{gray60}{rgb}{0.60,0.60,0.60}
\definecolor{gray61}{rgb}{0.61,0.61,0.61}
\definecolor{gray62}{rgb}{0.62,0.62,0.62}
\definecolor{gray63}{rgb}{0.63,0.63,0.63}
\definecolor{gray64}{rgb}{0.64,0.64,0.64}
\definecolor{gray65}{rgb}{0.65,0.65,0.65}
\definecolor{gray66}{rgb}{0.66,0.66,0.66}
\definecolor{gray67}{rgb}{0.67,0.67,0.67}
\definecolor{gray68}{rgb}{0.68,0.68,0.68}
\definecolor{gray69}{rgb}{0.69,0.69,0.69}
\definecolor{gray6}{rgb}{0.06,0.06,0.06}
\definecolor{gray70}{rgb}{0.70,0.70,0.70}
\definecolor{gray71}{rgb}{0.71,0.71,0.71}
\definecolor{gray72}{rgb}{0.72,0.72,0.72}
\definecolor{gray73}{rgb}{0.73,0.73,0.73}
\definecolor{gray74}{rgb}{0.74,0.74,0.74}
\definecolor{gray75}{rgb}{0.75,0.75,0.75}
\definecolor{gray76}{rgb}{0.76,0.76,0.76}
\definecolor{gray77}{rgb}{0.77,0.77,0.77}
\definecolor{gray78}{rgb}{0.78,0.78,0.78}
\definecolor{gray79}{rgb}{0.79,0.79,0.79}
\definecolor{gray7}{rgb}{0.07,0.07,0.07}
\definecolor{gray80}{rgb}{0.80,0.80,0.80}
\definecolor{gray81}{rgb}{0.81,0.81,0.81}
\definecolor{gray82}{rgb}{0.82,0.82,0.82}
\definecolor{gray83}{rgb}{0.83,0.83,0.83}
\definecolor{gray84}{rgb}{0.84,0.84,0.84}
\definecolor{gray85}{rgb}{0.85,0.85,0.85}
\definecolor{gray86}{rgb}{0.86,0.86,0.86}
\definecolor{gray87}{rgb}{0.87,0.87,0.87}
\definecolor{gray88}{rgb}{0.88,0.88,0.88}
\definecolor{gray89}{rgb}{0.89,0.89,0.89}
\definecolor{gray8}{rgb}{0.08,0.08,0.08}
\definecolor{gray90}{rgb}{0.90,0.90,0.90}
\definecolor{gray91}{rgb}{0.91,0.91,0.91}
\definecolor{gray92}{rgb}{0.92,0.92,0.92}
\definecolor{gray93}{rgb}{0.93,0.93,0.93}
\definecolor{gray94}{rgb}{0.94,0.94,0.94}
\definecolor{gray95}{rgb}{0.95,0.95,0.95}
\definecolor{gray96}{rgb}{0.96,0.96,0.96}
\definecolor{gray97}{rgb}{0.97,0.97,0.97}
\definecolor{gray98}{rgb}{0.98,0.98,0.98}
\definecolor{gray99}{rgb}{0.99,0.99,0.99}
\definecolor{gray9}{rgb}{0.09,0.09,0.09}
\definecolor{gray}{rgb}{0.75,0.75,0.75}
\definecolor{green1}{rgb}{0.00,1.00,0.00}
\definecolor{green2}{rgb}{0.00,0.93,0.00}
\definecolor{green3}{rgb}{0.00,0.80,0.00}
\definecolor{green4}{rgb}{0.00,0.55,0.00}
\definecolor{greenyellow}{rgb}{0.68,1.00,0.18}
\definecolor{green}{rgb}{0.00,1.00,0.00}
\definecolor{grey0}{rgb}{0.00,0.00,0.00}
\definecolor{grey100}{rgb}{1.00,1.00,1.00}
\definecolor{grey10}{rgb}{0.10,0.10,0.10}
\definecolor{grey11}{rgb}{0.11,0.11,0.11}
\definecolor{grey12}{rgb}{0.12,0.12,0.12}
\definecolor{grey13}{rgb}{0.13,0.13,0.13}
\definecolor{grey14}{rgb}{0.14,0.14,0.14}
\definecolor{grey15}{rgb}{0.15,0.15,0.15}
\definecolor{grey16}{rgb}{0.16,0.16,0.16}
\definecolor{grey17}{rgb}{0.17,0.17,0.17}
\definecolor{grey18}{rgb}{0.18,0.18,0.18}
\definecolor{grey19}{rgb}{0.19,0.19,0.19}
\definecolor{grey1}{rgb}{0.01,0.01,0.01}
\definecolor{grey20}{rgb}{0.20,0.20,0.20}
\definecolor{grey21}{rgb}{0.21,0.21,0.21}
\definecolor{grey22}{rgb}{0.22,0.22,0.22}
\definecolor{grey23}{rgb}{0.23,0.23,0.23}
\definecolor{grey24}{rgb}{0.24,0.24,0.24}
\definecolor{grey25}{rgb}{0.25,0.25,0.25}
\definecolor{grey26}{rgb}{0.26,0.26,0.26}
\definecolor{grey27}{rgb}{0.27,0.27,0.27}
\definecolor{grey28}{rgb}{0.28,0.28,0.28}
\definecolor{grey29}{rgb}{0.29,0.29,0.29}
\definecolor{grey2}{rgb}{0.02,0.02,0.02}
\definecolor{grey30}{rgb}{0.30,0.30,0.30}
\definecolor{grey31}{rgb}{0.31,0.31,0.31}
\definecolor{grey32}{rgb}{0.32,0.32,0.32}
\definecolor{grey33}{rgb}{0.33,0.33,0.33}
\definecolor{grey34}{rgb}{0.34,0.34,0.34}
\definecolor{grey35}{rgb}{0.35,0.35,0.35}
\definecolor{grey36}{rgb}{0.36,0.36,0.36}
\definecolor{grey37}{rgb}{0.37,0.37,0.37}
\definecolor{grey38}{rgb}{0.38,0.38,0.38}
\definecolor{grey39}{rgb}{0.39,0.39,0.39}
\definecolor{grey3}{rgb}{0.03,0.03,0.03}
\definecolor{grey40}{rgb}{0.40,0.40,0.40}
\definecolor{grey41}{rgb}{0.41,0.41,0.41}
\definecolor{grey42}{rgb}{0.42,0.42,0.42}
\definecolor{grey43}{rgb}{0.43,0.43,0.43}
\definecolor{grey44}{rgb}{0.44,0.44,0.44}
\definecolor{grey45}{rgb}{0.45,0.45,0.45}
\definecolor{grey46}{rgb}{0.46,0.46,0.46}
\definecolor{grey47}{rgb}{0.47,0.47,0.47}
\definecolor{grey48}{rgb}{0.48,0.48,0.48}
\definecolor{grey49}{rgb}{0.49,0.49,0.49}
\definecolor{grey4}{rgb}{0.04,0.04,0.04}
\definecolor{grey50}{rgb}{0.50,0.50,0.50}
\definecolor{grey51}{rgb}{0.51,0.51,0.51}
\definecolor{grey52}{rgb}{0.52,0.52,0.52}
\definecolor{grey53}{rgb}{0.53,0.53,0.53}
\definecolor{grey54}{rgb}{0.54,0.54,0.54}
\definecolor{grey55}{rgb}{0.55,0.55,0.55}
\definecolor{grey56}{rgb}{0.56,0.56,0.56}
\definecolor{grey57}{rgb}{0.57,0.57,0.57}
\definecolor{grey58}{rgb}{0.58,0.58,0.58}
\definecolor{grey59}{rgb}{0.59,0.59,0.59}
\definecolor{grey5}{rgb}{0.05,0.05,0.05}
\definecolor{grey60}{rgb}{0.60,0.60,0.60}
\definecolor{grey61}{rgb}{0.61,0.61,0.61}
\definecolor{grey62}{rgb}{0.62,0.62,0.62}
\definecolor{grey63}{rgb}{0.63,0.63,0.63}
\definecolor{grey64}{rgb}{0.64,0.64,0.64}
\definecolor{grey65}{rgb}{0.65,0.65,0.65}
\definecolor{grey66}{rgb}{0.66,0.66,0.66}
\definecolor{grey67}{rgb}{0.67,0.67,0.67}
\definecolor{grey68}{rgb}{0.68,0.68,0.68}
\definecolor{grey69}{rgb}{0.69,0.69,0.69}
\definecolor{grey6}{rgb}{0.06,0.06,0.06}
\definecolor{grey70}{rgb}{0.70,0.70,0.70}
\definecolor{grey71}{rgb}{0.71,0.71,0.71}
\definecolor{grey72}{rgb}{0.72,0.72,0.72}
\definecolor{grey73}{rgb}{0.73,0.73,0.73}
\definecolor{grey74}{rgb}{0.74,0.74,0.74}
\definecolor{grey75}{rgb}{0.75,0.75,0.75}
\definecolor{grey76}{rgb}{0.76,0.76,0.76}
\definecolor{grey77}{rgb}{0.77,0.77,0.77}
\definecolor{grey78}{rgb}{0.78,0.78,0.78}
\definecolor{grey79}{rgb}{0.79,0.79,0.79}
\definecolor{grey7}{rgb}{0.07,0.07,0.07}
\definecolor{grey80}{rgb}{0.80,0.80,0.80}
\definecolor{grey81}{rgb}{0.81,0.81,0.81}
\definecolor{grey82}{rgb}{0.82,0.82,0.82}
\definecolor{grey83}{rgb}{0.83,0.83,0.83}
\definecolor{grey84}{rgb}{0.84,0.84,0.84}
\definecolor{grey85}{rgb}{0.85,0.85,0.85}
\definecolor{grey86}{rgb}{0.86,0.86,0.86}
\definecolor{grey87}{rgb}{0.87,0.87,0.87}
\definecolor{grey88}{rgb}{0.88,0.88,0.88}
\definecolor{grey89}{rgb}{0.89,0.89,0.89}
\definecolor{grey8}{rgb}{0.08,0.08,0.08}
\definecolor{grey90}{rgb}{0.90,0.90,0.90}
\definecolor{grey91}{rgb}{0.91,0.91,0.91}
\definecolor{grey92}{rgb}{0.92,0.92,0.92}
\definecolor{grey93}{rgb}{0.93,0.93,0.93}
\definecolor{grey94}{rgb}{0.94,0.94,0.94}
\definecolor{grey95}{rgb}{0.95,0.95,0.95}
\definecolor{grey96}{rgb}{0.96,0.96,0.96}
\definecolor{grey97}{rgb}{0.97,0.97,0.97}
\definecolor{grey98}{rgb}{0.98,0.98,0.98}
\definecolor{grey99}{rgb}{0.99,0.99,0.99}
\definecolor{grey9}{rgb}{0.09,0.09,0.09}
\definecolor{grey}{rgb}{0.75,0.75,0.75}
\definecolor{honeydew1}{rgb}{0.94,1.00,0.94}
\definecolor{honeydew2}{rgb}{0.88,0.93,0.88}
\definecolor{honeydew3}{rgb}{0.76,0.80,0.76}
\definecolor{honeydew4}{rgb}{0.51,0.55,0.51}
\definecolor{honeydew}{rgb}{0.94,1.00,0.94}
\definecolor{hotpink}{rgb}{1.00,0.41,0.71}
\definecolor{indianred}{rgb}{0.80,0.36,0.36}
\definecolor{ivory1}{rgb}{1.00,1.00,0.94}
\definecolor{ivory2}{rgb}{0.93,0.93,0.88}
\definecolor{ivory3}{rgb}{0.80,0.80,0.76}
\definecolor{ivory4}{rgb}{0.55,0.55,0.51}
\definecolor{ivory}{rgb}{1.00,1.00,0.94}
\definecolor{khaki1}{rgb}{1.00,0.96,0.56}
\definecolor{khaki2}{rgb}{0.93,0.90,0.52}
\definecolor{khaki3}{rgb}{0.80,0.78,0.45}
\definecolor{khaki4}{rgb}{0.55,0.53,0.31}
\definecolor{khaki}{rgb}{0.94,0.90,0.55}
\definecolor{lavenderblush}{rgb}{1.00,0.94,0.96}
\definecolor{lavender}{rgb}{0.90,0.90,0.98}
\definecolor{lawngreen}{rgb}{0.49,0.99,0.00}
\definecolor{lemonchiffon}{rgb}{1.00,0.98,0.80}
\definecolor{lightblue}{rgb}{0.68,0.85,0.90}
\definecolor{lightcoral}{rgb}{0.94,0.50,0.50}
\definecolor{lightcyan}{rgb}{0.88,1.00,1.00}
\definecolor{lightgoldenrod}{rgb}{0.93,0.87,0.51}
\definecolor{lightgoldenrod}{rgb}{0.98,0.98,0.82}
\definecolor{lightgray}{rgb}{0.83,0.83,0.83}
\definecolor{lightgreen}{rgb}{0.56,0.93,0.56}
\definecolor{lightgrey}{rgb}{0.83,0.83,0.83}
\definecolor{lightpink}{rgb}{1.00,0.71,0.76}
\definecolor{lightsalmon}{rgb}{1.00,0.63,0.48}
\definecolor{lightsea}{rgb}{0.13,0.70,0.67}
\definecolor{lightsky}{rgb}{0.53,0.81,0.98}
\definecolor{lightslate}{rgb}{0.47,0.53,0.60}
\definecolor{lightslate}{rgb}{0.47,0.53,0.60}
\definecolor{lightslate}{rgb}{0.52,0.44,1.00}
\definecolor{lightsteel}{rgb}{0.69,0.77,0.87}
\definecolor{lightyellow}{rgb}{1.00,1.00,0.88}
\definecolor{limegreen}{rgb}{0.20,0.80,0.20}
\definecolor{linen}{rgb}{0.98,0.94,0.90}
\definecolor{magenta1}{rgb}{1.00,0.00,1.00}
\definecolor{magenta2}{rgb}{0.93,0.00,0.93}
\definecolor{magenta3}{rgb}{0.80,0.00,0.80}
\definecolor{magenta4}{rgb}{0.55,0.00,0.55}
\definecolor{magenta}{rgb}{1.00,0.00,1.00}
\definecolor{maroon1}{rgb}{1.00,0.20,0.70}
\definecolor{maroon2}{rgb}{0.93,0.19,0.65}
\definecolor{maroon3}{rgb}{0.80,0.16,0.56}
\definecolor{maroon4}{rgb}{0.55,0.11,0.38}
\definecolor{maroon}{rgb}{0.69,0.19,0.38}
\definecolor{mediumaquamarine}{rgb}{0.40,0.80,0.67}
\definecolor{mediumblue}{rgb}{0.00,0.00,0.80}
\definecolor{mediumorchid}{rgb}{0.73,0.33,0.83}
\definecolor{mediumpurple}{rgb}{0.58,0.44,0.86}
\definecolor{mediumsea}{rgb}{0.24,0.70,0.44}
\definecolor{mediumslate}{rgb}{0.48,0.41,0.93}
\definecolor{mediumspring}{rgb}{0.00,0.98,0.60}
\definecolor{mediumturquoise}{rgb}{0.28,0.82,0.80}
\definecolor{mediumviolet}{rgb}{0.78,0.08,0.52}
\definecolor{midnightblue}{rgb}{0.10,0.10,0.44}
\definecolor{mintcream}{rgb}{0.96,1.00,0.98}
\definecolor{mistyrose}{rgb}{1.00,0.89,0.88}
\definecolor{moccasin}{rgb}{1.00,0.89,0.71}
\definecolor{navajowhite}{rgb}{1.00,0.87,0.68}
\definecolor{navyblue}{rgb}{0.00,0.00,0.50}
\definecolor{navy}{rgb}{0.00,0.00,0.50}
\definecolor{oldlace}{rgb}{0.99,0.96,0.90}
\definecolor{olivedrab}{rgb}{0.42,0.56,0.14}
\definecolor{orange1}{rgb}{1.00,0.65,0.00}
\definecolor{orange2}{rgb}{0.93,0.60,0.00}
\definecolor{orange3}{rgb}{0.80,0.52,0.00}
\definecolor{orange4}{rgb}{0.55,0.35,0.00}
\definecolor{orangered}{rgb}{1.00,0.27,0.00}
\definecolor{orange}{rgb}{1.00,0.65,0.00}
\definecolor{orchid1}{rgb}{1.00,0.51,0.98}
\definecolor{orchid2}{rgb}{0.93,0.48,0.91}
\definecolor{orchid3}{rgb}{0.80,0.41,0.79}
\definecolor{orchid4}{rgb}{0.55,0.28,0.54}
\definecolor{orchid}{rgb}{0.85,0.44,0.84}
\definecolor{palegoldenrod}{rgb}{0.93,0.91,0.67}
\definecolor{palegreen}{rgb}{0.60,0.98,0.60}
\definecolor{paleturquoise}{rgb}{0.69,0.93,0.93}
\definecolor{paleviolet}{rgb}{0.86,0.44,0.58}
\definecolor{papayawhip}{rgb}{1.00,0.94,0.84}
\definecolor{peachpuff}{rgb}{1.00,0.85,0.73}
\definecolor{peru}{rgb}{0.80,0.52,0.25}
\definecolor{pink1}{rgb}{1.00,0.71,0.77}
\definecolor{pink2}{rgb}{0.93,0.66,0.72}
\definecolor{pink3}{rgb}{0.80,0.57,0.62}
\definecolor{pink4}{rgb}{0.55,0.39,0.42}
\definecolor{pink}{rgb}{1.00,0.75,0.80}
\definecolor{plum1}{rgb}{1.00,0.73,1.00}
\definecolor{plum2}{rgb}{0.93,0.68,0.93}
\definecolor{plum3}{rgb}{0.80,0.59,0.80}
\definecolor{plum4}{rgb}{0.55,0.40,0.55}
\definecolor{plum}{rgb}{0.87,0.63,0.87}
\definecolor{powderblue}{rgb}{0.69,0.88,0.90}
\definecolor{purple1}{rgb}{0.61,0.19,1.00}
\definecolor{purple2}{rgb}{0.57,0.17,0.93}
\definecolor{purple3}{rgb}{0.49,0.15,0.80}
\definecolor{purple4}{rgb}{0.33,0.10,0.55}
\definecolor{purple}{rgb}{0.63,0.13,0.94}
\definecolor{red1}{rgb}{1.00,0.00,0.00}
\definecolor{red2}{rgb}{0.93,0.00,0.00}
\definecolor{red3}{rgb}{0.80,0.00,0.00}
\definecolor{red4}{rgb}{0.55,0.00,0.00}
\definecolor{red}{rgb}{1.00,0.00,0.00}
\definecolor{rosybrown}{rgb}{0.74,0.56,0.56}
\definecolor{royalblue}{rgb}{0.25,0.41,0.88}
\definecolor{saddlebrown}{rgb}{0.55,0.27,0.07}
\definecolor{salmon1}{rgb}{1.00,0.55,0.41}
\definecolor{salmon2}{rgb}{0.93,0.51,0.38}
\definecolor{salmon3}{rgb}{0.80,0.44,0.33}
\definecolor{salmon4}{rgb}{0.55,0.30,0.22}
\definecolor{salmon}{rgb}{0.98,0.50,0.45}
\definecolor{sandybrown}{rgb}{0.96,0.64,0.38}
\definecolor{seagreen}{rgb}{0.18,0.55,0.34}
\definecolor{seashell1}{rgb}{1.00,0.96,0.93}
\definecolor{seashell2}{rgb}{0.93,0.90,0.87}
\definecolor{seashell3}{rgb}{0.80,0.77,0.75}
\definecolor{seashell4}{rgb}{0.55,0.53,0.51}
\definecolor{seashell}{rgb}{1.00,0.96,0.93}
\definecolor{sienna1}{rgb}{1.00,0.51,0.28}
\definecolor{sienna2}{rgb}{0.93,0.47,0.26}
\definecolor{sienna3}{rgb}{0.80,0.41,0.22}
\definecolor{sienna4}{rgb}{0.55,0.28,0.15}
\definecolor{sienna}{rgb}{0.63,0.32,0.18}
\definecolor{skyblue}{rgb}{0.53,0.81,0.92}
\definecolor{slateblue}{rgb}{0.42,0.35,0.80}
\definecolor{slategray}{rgb}{0.44,0.50,0.56}
\definecolor{slategrey}{rgb}{0.44,0.50,0.56}
\definecolor{snow1}{rgb}{1.00,0.98,0.98}
\definecolor{snow2}{rgb}{0.93,0.91,0.91}
\definecolor{snow3}{rgb}{0.80,0.79,0.79}
\definecolor{snow4}{rgb}{0.55,0.54,0.54}
\definecolor{snow}{rgb}{1.00,0.98,0.98}
\definecolor{springgreen}{rgb}{0.00,1.00,0.50}
\definecolor{steelblue}{rgb}{0.27,0.51,0.71}
\definecolor{tan1}{rgb}{1.00,0.65,0.31}
\definecolor{tan2}{rgb}{0.93,0.60,0.29}
\definecolor{tan3}{rgb}{0.80,0.52,0.25}
\definecolor{tan4}{rgb}{0.55,0.35,0.17}
\definecolor{tan}{rgb}{0.82,0.71,0.55}
\definecolor{thistle1}{rgb}{1.00,0.88,1.00}
\definecolor{thistle2}{rgb}{0.93,0.82,0.93}
\definecolor{thistle3}{rgb}{0.80,0.71,0.80}
\definecolor{thistle4}{rgb}{0.55,0.48,0.55}
\definecolor{thistle}{rgb}{0.85,0.75,0.85}
\definecolor{tomato1}{rgb}{1.00,0.39,0.28}
\definecolor{tomato2}{rgb}{0.93,0.36,0.26}
\definecolor{tomato3}{rgb}{0.80,0.31,0.22}
\definecolor{tomato4}{rgb}{0.55,0.21,0.15}
\definecolor{tomato}{rgb}{1.00,0.39,0.28}
\definecolor{turquoise1}{rgb}{0.00,0.96,1.00}
\definecolor{turquoise2}{rgb}{0.00,0.90,0.93}
\definecolor{turquoise3}{rgb}{0.00,0.77,0.80}
\definecolor{turquoise4}{rgb}{0.00,0.53,0.55}
\definecolor{turquoise}{rgb}{0.25,0.88,0.82}
\definecolor{violetred}{rgb}{0.82,0.13,0.56}
\definecolor{violet}{rgb}{0.93,0.51,0.93}
\definecolor{wheat1}{rgb}{1.00,0.91,0.73}
\definecolor{wheat2}{rgb}{0.93,0.85,0.68}
\definecolor{wheat3}{rgb}{0.80,0.73,0.59}
\definecolor{wheat4}{rgb}{0.55,0.49,0.40}
\definecolor{wheat}{rgb}{0.96,0.87,0.70}
\definecolor{whitesmoke}{rgb}{0.96,0.96,0.96}
\definecolor{white}{rgb}{1.00,1.00,1.00}
\definecolor{yellow1}{rgb}{1.00,1.00,0.00}
\definecolor{yellow2}{rgb}{0.93,0.93,0.00}
\definecolor{yellow3}{rgb}{0.80,0.80,0.00}
\definecolor{yellow4}{rgb}{0.55,0.55,0.00}
\definecolor{yellowgreen}{rgb}{0.60,0.80,0.20}
\definecolor{yellow}{rgb}{1.00,1.00,0.00}

\usepackage{graphicx}
\usepackage{dcolumn}
\usepackage{bm}
\usepackage{hyperref}
\usepackage{ulem}

\begin{document}


\title{Crystallization and order-disorder transition of colloidal particles in a drying suspension: a phase field crystal approach}
\author{Nirmalendu Ganai$^{1}$, Arnab Saha$^{2}$, Surajit Sengupta$^{3,4}$}
\address{$^1$ Department of Physics, Nabadwip Vidyasagar College, Nabadwip, Nadia $741302$, India\\$^2$ Max Planck Institute for the Physics of Complex Systems, N{\"o}thnitzer Stra{\ss}e $38$, $01187$ Dresden, Germany\\$^{3}$TIFR Centre for Interdisciplinary Sciences, $21$ Brundavan Colony, Narsingi, Hyderabad $500075$, India\\$^{4}$Centre for Advanced Materials, Indian Association for the Cultivation of Science, Jadavpur, Kolkata $700032$ , India}

\ead{surajit{\_}tcis@tifr.res.in}

\date{\today}

\begin{abstract}
Using a phase field crystal model we study the structure and dynamics of a drop of colloidal suspension during evaporation of the solvent. We model an experimental system where contact line pinning of the drop on the substrate is non-existent. Under such carefully controlled conditions,  evaporation of the drop produces an ordered or disordered arrangement of the colloidal residue depending on the initial average density of solute and the drying rate. We obtain a non-equilibrium phase boundary showing amorphous and crystalline phases of single component and binary mixtures of colloidal particles in the density-drying rate plane. While single component colloids order in the two dimensional triangular lattice, a symmetric binary mixture of mutually repulsive particles orders in a three sub-lattice order where two of the sub-lattices of the triangular lattice are occupied by the two species of particles with the third sub-lattice vacant.   

\end{abstract}

\section{Introduction} The phase field crystal (PFC) technique has been used extensively to study a variety of phenomena like solidification, dendrite formation in binary alloys \cite{berry}, order-disorder transitions, and the elastic \cite{{grant},{haataja}} and rheological response of solids under external loads. The technique offers the unique advantage of being able to model atomistic detail within a coarse-grained, free energy functional approach. While there have been attempts to relate PFC \cite{{grant},{lowen}} models to more realistic dynamical density functional theory \cite{{ramakrishnan},{evans},{marconi}} approaches, the basic simplicity of PFC is appealing.

In this paper we present our results on the application of the PFC model in its simplest version to a complex problem of great technological interest viz. assembly of nano-particles from a drying suspension \cite{{col-assembl},{col-assembl-2},{non-eq}}. The experiment consists of placing a drop of suspension containing colloidal particles on a substrate and allowing the solvent to evaporate \cite{DD}. Under suitable conditions, which we explain shortly, the residue may consist of particles arranged in a regular periodic order. Such regular arrangements of nano-particles have many applications from functional arrays of sensors to templates for the production of more complicated nano-structures. 

The physics of the drying process is quite complicated and may be briefly described as follows. The rate of evaporation at the contact line, where the drop meets the substrate, is much higher than that of at the middle of the drop due to higher curvature \cite{Nagel}.  So, the evaporation of the drop proceeds mainly from the contact line \cite{nagayama}. As the solvent dries, a radially outward flow of solvent is set up which replenishes the evaporated liquid \cite{{evap-approx},{marangoni},{degan}} and tends to advect particles from the center to the periphery. Simultaneously the contact line moves inward towards the center of the drop with evaporation, dragging particles along with it due to surface tension forces. These competing flows result in a buildup of solute at the contact line which effectively pins it. Subsequently, with the stationary contact line, fluid evaporation changes contact angle \cite{jpcb} such that the pinning force decreases suddenly and the contact line becomes unpinned starting the cycle again. Such repeated pinning-de pinning transitions leave concentric rings of solute in the wake of the drying drop leading to the so-called ``coffee ring'' effect. Inhomogeneities of the solute density produced by this process leads to a disordered residue with the fluid having to flow through a highly ramified network of stationary and moving particles. Analysis of this complete process is complex and to our knowledge a complete solution to this problem does not exist. Also the disordered residue patterns produced makes it impossible to use this process as an effective technique of nano-particle assembly. 

From the above discussion, it is clear that suppressing contact line pinning would result in a drastic reduction of complexity and has the promise of producing an ordered residue \cite{{degan},{nagayama1}}. Indeed, recent experiments on the drying of colloidal suspensions on carefully prepared substrates do show such drying induced ordering \cite{{DD},{evap1},{evap2}}. A mobile contact line which moves uniformly leads to a steadily shrinking drop where solute particles experience only an inward compressing force arising from the moving meniscus. The radial compaction leads to solidification once the density surpasses the freezing density. However, if the drying rate is too large, the liquid evaporates before this density is reached and the colloidal particles are stranded in a disordered arrangement.

We have, in the past, modeled this system of uniformly drying colloidal suspension on a substrate without contact line pinning using Brownian dynamics simulations of particles interacting with a coarse-grained solvent field \cite{DD}. Here we use a PFC approach to simulate the system using two coupled coarse-grained fields corresponding to the solute and the solvent local densities respectively. Preliminary results of this study has been published as a conference proceedings \cite{ganai}. Finally, we have also generalized our system to represent a symmetric binary mixture of particles and we consider the case where the particles repel each other. In this case, there are three coupled coarse-grained fields representing the local densities of the two species of particles and the solvent. 

The paper is organized as follows. In the next section, we introduce our PFC model for single component suspension. This is followed by details of the numerical method used to solve the resulting coupled dynamical equations. In section 4, we present our results for the single component system. We generalize to the binary system in section 5 and present the results. Finally we end the paper with a brief discussion of our results and pointing out future directions of further work. 

\section{Phase field crystal model for a single component suspension} We begin by considering the evaporation of the solvent as a two dimensional (2d) liquid-gas phase transition. The phase separating fluid is described by a free energy functional ${{\cal F}_1}$ containing quadratic and quartic terms and an external chemical potential as follows:

\begin{equation}
{\cal F}_1[ \lbrace \phi \rbrace ] = \int d^2{\bf r}  \,\, [-a \phi^2 + b \phi^4 + c (\nabla \phi)^2 + h \phi],
\label{b}
\end{equation}

\noindent
Here $\phi({{\bf} r},t)$ is a scalar order parameter field (local density of the solvent) at position ${{\bf}r}$ and time $t$ with $a,b$ and $c > 0$ as phenomenological parameters. The free energy functional here is the standard `$\phi^4$' model of `liquid-gas' phase transition with the external field `$h$', representing a chemical potential which favors $\phi < 0$ (vapour) regions i.e., drives the evaporation process. The time-evolution of the fluid is determined by the time dependent Ginzburg-Landau (model A) dynamics\cite{chaikin},

\begin{eqnarray}
\dot \phi &=& - \Gamma \frac{\delta {\cal F}_1}{\delta \phi} \nonumber \\
&=& \Gamma (2a \phi - 4b \phi^3 +2c (\nabla^2 \phi) - h)
\label{a} 
\end{eqnarray} 
 
\noindent 
 The drying rate of the solvent can be controlled by varying the parameters $h$ and $\Gamma$, the kinetic coefficient. For small values of $h$ the free energy density in Eq.\ref{b} has minima at $\phi_{1}$ and $\phi_{2}$ where $\phi_{1}>0$ and $\phi_{2}<0$. We denote $\phi=\phi_{1}$ as liquid phase and $\phi=\phi_{2}$ as vapor phase of the solvent. For $h > 0$,  the solvent starts to evaporate and when all the solvent dries up then $\phi$ becomes $\phi_2$ everywhere.
 
\begin{figure}[h]
\begin{center}
\includegraphics[width=.3\textwidth]{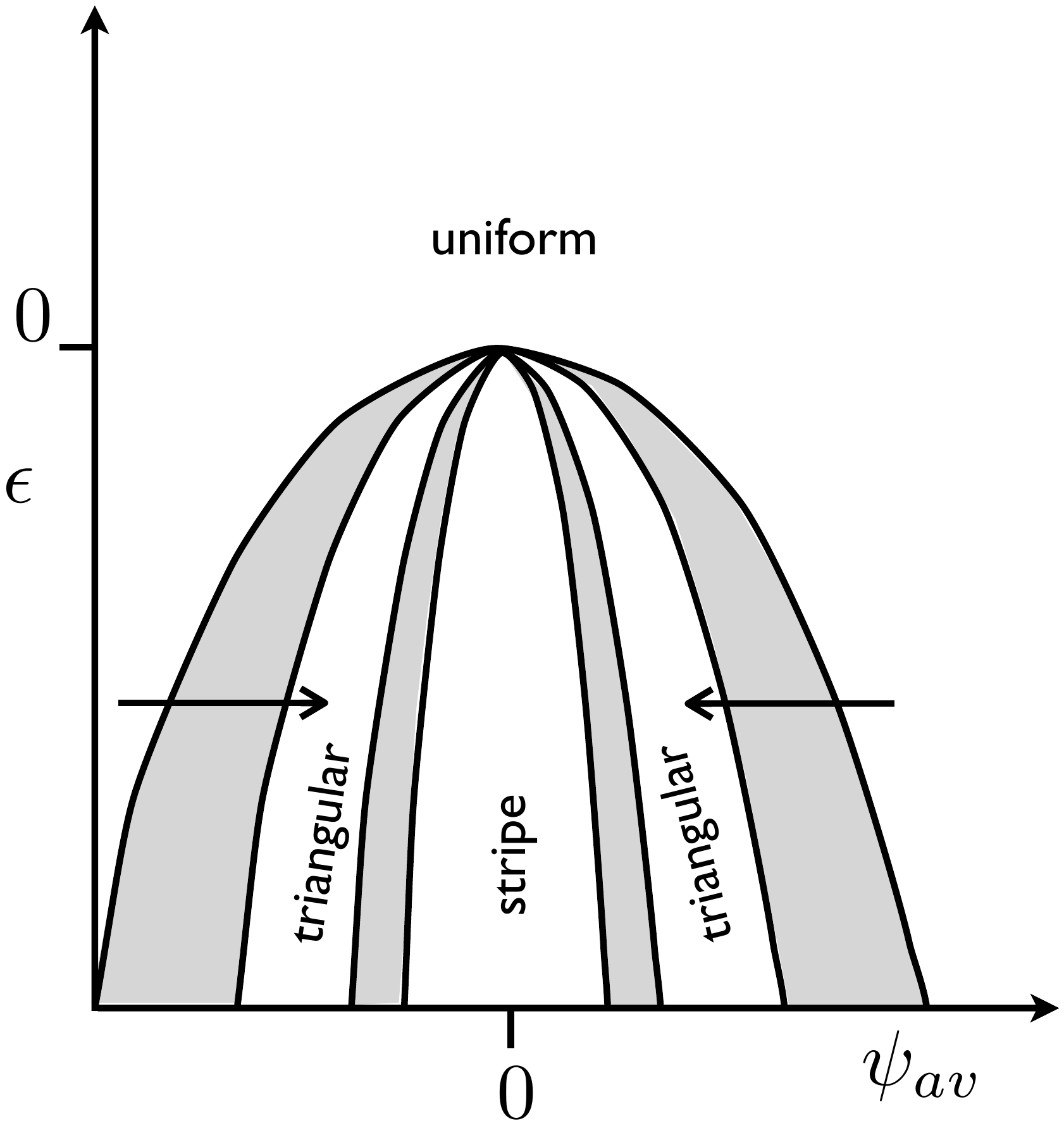}
\caption{ A schematic of the phase diagram in the $\psi_{av}-\epsilon$ plane from Ref.\cite{grant} obtained by minimizing the the Landau free energy functional ${\cal F}_2$ (Eq. \ref{cem}) over choices of functions $\psi ({\bf r})$. Note that the phase diagram is symmetric along $\psi_{av} = 0$ reflecting the basic symmetry of the functional. There are three distinct phases where (1) $\psi$ is uniform as in a liquid, (2) is periodic only in one direction (smectic or stripe phase) and (3) has the periodicity of the two dimensional triangular lattice. The shaded areas are coexistence regions and the arrows show quench paths connecting liquid and triangular solid phases.}
\label{pdia}
\end{center} 
\end{figure}

For solute particles we consider a PFC model in its simplest form defined by a dimensionless free energy functional \cite{grant,haataja} is

\begin{equation}
{\cal F}_2[\{\psi\}] = \int d^2{\bf r} \,\, \Big \lbrace \frac {\psi}{2} [\epsilon + (1 + \nabla^2)^2] \psi + \frac {\psi^4}{4} \Big \rbrace
\label{cem}
\end{equation}

\noindent
where $\psi$ is a coarse-grained field whose extrema corresponds to the equilibrium positions of the atoms and therefore may be interpreted as a dimensionless density field \cite{grant} of the solute. The phase diagram of this system, obtained by minimizing Eq.\ref{cem}, has been extensively studied in Ref.\cite{grant}. A schematic of the phase diagram obtained in Ref.\cite{grant} is reproduced in Fig.\ref{pdia}. The stable phase for all $\epsilon > 0$ is the uniform phase where $\psi = {\rm constant} = \psi_{av}$, where $\psi_{av}$ is the spatial average of $\psi$. As $\epsilon$ becomes negative, the free energy is minimized by modulated phases with modulations in either one or two dimensions. Further, the phase diagram in the $\psi_{av} - \epsilon$ plane is symmetric in $\psi_{av}$. 

There are some inherent difficulties involved in using the PFC model to describe the drying induced ordering of colloids. Firstly, there is no evidence for a smectic phase in a colloidal system with spherical particles, so this phase is unphysical and needs to be discarded. Secondly, unlike the usual freezing phase diagram where a solid is obtained from the liquid on cooling or increasing the density, in the PFC model, the transition is re-entrant so that there are limits to how much $\psi_{av}$ can be increased at fixed $\epsilon$ such that one obtains a periodic solid from the uniform phase. Finally, the order parameter jump across the freezing transition can become very small for small $|\epsilon|$, this is in variance with a real liquid - solid transition which is strongly first order and remains over all physically accessible temperatures and pressures for most common substances.  In spite of these caveats, it is possible to work within a range of parameters such that meaningful results are obtained as we describe below.


 In our case as the solvent starts to evaporate continually from the contact line, the liquid-gas interface shrinks and consequently the solute particles are forced to come together along the radially inward direction and hence, the value of $\psi_{av}$ increases. The evaporation of the solute therefore induces a force on the particles coupling the fields $\phi$ and $\psi$ in the dimensionless free energy ${\cal F}_2[\{\psi\}]$. Rewriting  ${\cal F}_2[\{\psi\}]$ including an appropriate (leading order) coupling gives us 
 
 \begin{equation}
{\cal F}_2 = \int d^2{\bf r} \,\, \Big \lbrace \frac {\psi}{2} [\epsilon + (1 + \nabla^2)^2] \psi + \frac {\psi^4}{4} + \phi \psi \Big \rbrace
\label{exact}
\end{equation}
where the coupling coefficient has been chosen to be unity. 

The equation of motion for conserved field $\psi$ (using model B dynamics \cite{{berry},{chaikin}}) is therefore given by,

\begin{equation}
\frac {\partial \psi}{\partial t} =  \overrightarrow{\nabla} {\bf .} (\mu \overrightarrow{\nabla} \frac {\delta {\cal F}_2}{\delta \psi}) = \overrightarrow{\nabla} {\bf .}  \Big \lbrace \mu \overrightarrow{\nabla} \big ( [\epsilon + (1 + \nabla^2)^2] \psi + \psi^3 + \phi \big ) \Big \rbrace
\label{dynamic}
\end{equation}

In the above equation $\mu$ is the mobility of the solute which is governed by the following equation.

\begin{equation}
\mu = (\tanh(\gamma \phi)+1)/2
\label{mueq}
\end{equation}

The functional form of $\mu$ should encode the fact that if solute particles leave the solvent during evaporation, their motion ceases due to the lack of Brownian fluctuations in gas where $\phi=\phi_2 < 0$.  So we choose the functional form of $\mu$ such that as $\phi=\phi_2$, $\mu \to 0$. The value of $\gamma$ controls the rate at which $\mu$ vanishes in the limit $\phi \to \phi_2$, near the liquid-vapour interface. Here the average density of the solute particles and the drying rate are the control parameters. The average density of the solute particles can be varied by varying $\psi_{av}$. The drying rate can be controlled by varying $h$ and $\Gamma$ both.

As the liquid-vapour interface shrinks continuously enhancing $\psi_{av}$, it eventually reaches a range where a periodic $\psi$ is expected and  an ordered (triangular) phase forms (see Fig.\ref{pdia}).  On the other hand, if the drying rate and the initial value of $\psi_{av}$ is such that the final value of $\psi_{av}$ remains below the critical value needed for solidification till the final stage of evaporation, then a disordered solute phase is expected. 

In this model two important time scales are responsible for the `drying induced ordering' phenomena. The first, $\tau_1$,  is related to the drying time of the solvent and other, $\tau_2$, is related to the relaxation of the solute particles to its crystalline minima.  No ordering in solute particle is possible if $\psi_{av}$ is lower than the critical value and $\tau_1 < \tau_2$, because the solute particle does not get the sufficient time to reach the crystalline minima in that case. Ordering is only possible if $\tau_1$ $\geq$ $\tau_2$ and the $\psi_{av}$ is greater than the critical value.

We now give below details of the procedure used to solve the dynamics for the model given in Eqs.\ref{a} and \ref{dynamic}. The coupled dynamics of the solute and the solvent are not exactly solvable by analytical methods in general. However, the dynamics of a drying drop of pure solvent without solute {\it is} solvable using a reasonable ansatz for the shape of the drop as we show below. We utilize this fact to define two methods of solution. In method I \cite{ganai}, we solve for the dynamics of the drop radius analytically, which is then used as an input to solve the dynamics of the particles, i.e. $\psi$. In method II, the dynamics for both the coupled fields $\phi$ and $\psi$ are solved numerically. Both methods gives similar results but method I is faster than method II. Below we will describe both in order.      

In method I \cite{ganai}, for a given  $\psi_{av}$, we use an ansatz for $\phi$ remembering that it represents a two dimensional drop of solvent;

\begin{eqnarray}
\phi ({\bf r},t=0) &=& \phi_1 \,\,\,\,\,\,\, for \,\, |{\bf r}| = R \nonumber \\
&=& \phi_1 + \frac {(\phi_1 - \phi_2)}{\xi} (R - {\bf r}) \,\,\,\,\, for \,\, R \le  |{\bf r}|  \le (R + \xi) \nonumber \\
&=& \phi_2 \,\,\,\,\,\,\,  otherwise 
\label{radius}
\end{eqnarray}

Substituting this profile in the free energy Eq.\ref{a}, and minimizing it with respect to the interfacial width, we obtain $\xi = \sqrt[2]{\frac{c (\phi_1 - \phi_2)^3}{{\cal I}_s}}$ for $\frac {\xi}{R} \ll 1$, with ${{\cal I}_s}$ = $|\frac{-a(\phi_1^3 - \phi_2^3)}{3}+\frac{b(\phi_1^5 - \phi_2^5)}{5}+\frac{h(\phi_1^2 - \phi_2^2)}{2}|$. The rate of change of $\phi ({\bf r},t)$, $\frac {d \phi}{d t}$, is then obtained as,

\begin{equation}
\frac {d \phi}{d t} = \frac {d R}{d t} \,\, \frac {d \phi}{d {\bf r}} \Big |_{{\bf r}=R}
\label{radius1}
\end{equation}

which, after some straightforward algebra, gives the dynamical equation for the radius of the drop during evaporation as

\begin{eqnarray}
\frac {dR}{dt} &=& \dot R = \frac {\xi}{(\phi_1 - \phi_2)^2} [-a (\phi_1^2 - \phi_2^2) + b (\phi_1^4 - \phi_2^4) \nonumber \\
&& +\: h (\phi_1 - \phi_2)] + \frac {2 c}{R}
\label{approx}
\end{eqnarray}

Where we have assumed that evaporation depends only on radial coordinate such that the shape of the drop is kept fixed and $\frac {\xi}{R} \ll 1$. After integrating Eq.\ref{approx}, we get
\begin{equation}
{R} = {R_0} - \frac {2 c}{D} \ln (\frac {D {R_0} + 2 c}{DR + 2 c}) + Dt
\label{radius3}
\end{equation}
where the $R_0$ is the initial radius of the drop and $D = \frac {\xi}{(\phi_1 - \phi_2)^2} [-a (\phi_1^2 - \phi_2^2) + b (\phi_1^4 - \phi_2^4) + h (\phi_1 - \phi_2)]$.

\begin{figure}[h]
\begin{center}
\includegraphics[width=1.0\textwidth]{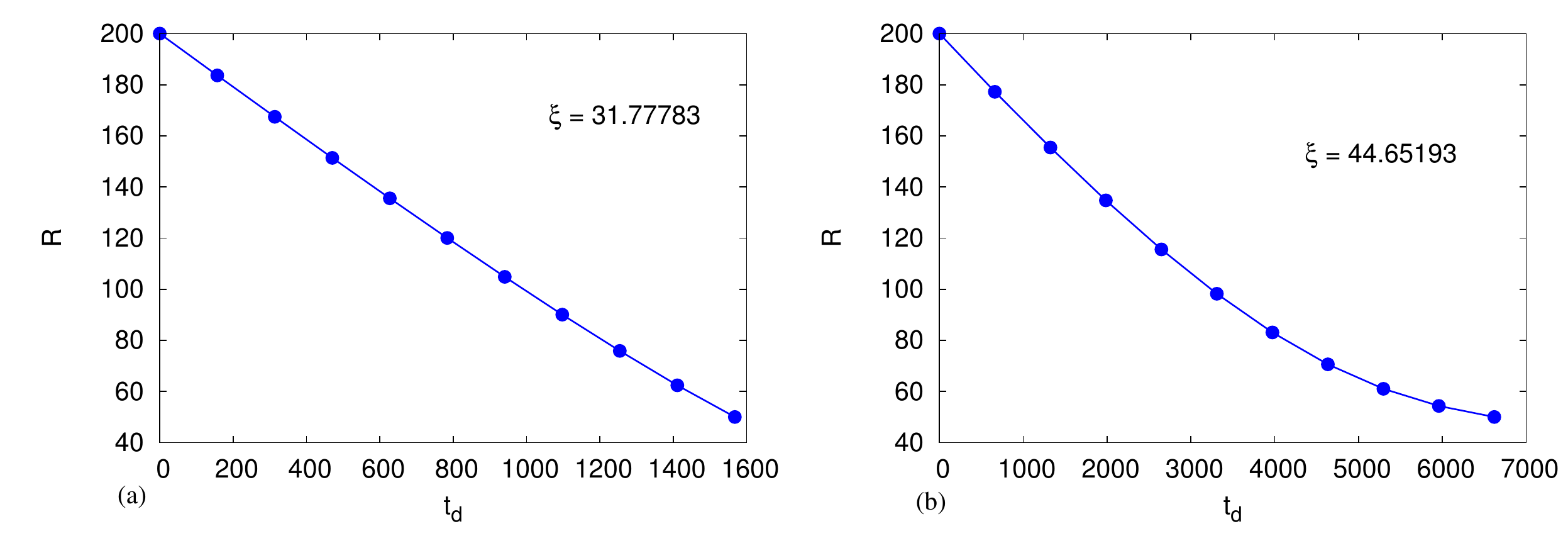}
\caption{ (a) and (b) show the variation of the radius of the solvent $(R)$ with time for $t_d = 1568.15, h=0.006$ (fast drying) and $t_d = 6620.32, h=0.002$ (slow drying) respectively. $R^* = 50.0$ in the both cases. Fig. is taken from \cite{ganai}.}
\label{Fig1}
\end{center} 
\end{figure}

Since Eq.\ref{radius3} is a transcendental equation, we solve the Eq.\ref{radius3} numerically using the bisection method at every time step. Fig.\ref{Fig1} shows typical solutions of $R(t)$. In this simulation we choose $\psi ({\bf r},t=0)$ randomly \cite{press} over space with a non-zero average value. At every time step we get the value of the radius of the drop solving Eq.\ref{radius3} numerically and with this value of radius a new configuration of solvent density $\phi({\bf r},t)$ is generated according to Eq.\ref{radius} at every time step. Using this value of $\phi({\bf r},t)$, we update $\psi({\bf r},t)$ using  Eq. \ref{dynamic}. A Fast Fourier Transformation along with Euler discretisation is used to solve Eq.\ref{dynamic}. When the value of the radius of the drop becomes equal to $R^*$ such that $\xi<R^*<< R_0$, we get the final configuration of $\psi$ and the time required to arrive at this stage, starting from $R_0$ is the total drying time ($t_d$). 

Method II, on the other hand, is more straight-forward, where we solve for both solvent as well as solute dynamics numerically, again using an explicit Euler discretization scheme with a suitably small time step.  At each time step we update the density profile of the solvent and use the updated profile to solve the solute-dynamics as before.
  
Finally, we fix the length and time scales for simulation and choose a set of values for the parameters of the model as follows. The length scale is set by the coefficient of the squared gradient term in the free energy for $\psi$ and the time scale is set by the coefficient of the time derivative of $\psi$, both of which are set to unity. We take $\gamma=10.0$ in Eq.\ref{mueq} to set the mobility of the solute particles inside the liquid as well as outside the liquid. The parameters $a=0.008,b=0.004,c=1.0$.  The value of $h$ is adjusted to control the drying time. Using these values, i.e., values of $a,b,c$, we get the minima of the dimensionless free energy ${\cal {F}}_1$ at $\phi_1=0.651$, $\phi_2=-1.15$ for $h=0.006$ (a typical value of $h$ that is used here for simulating fast drying)  and $\phi_1=0.930$, $\phi_2=-1.06$ for $h=0.002$( a typical value of $h$ that is used here for simulating slow drying). The interfacial width becomes, $\xi=31.77783$, for $h=0.006$ and $\xi=44.65193$, for $h=0.002$. A system of size $512 \Delta x \times 512 \Delta x$ with $\Delta x=1.0$ is taken with periodic boundary conditions. Time step $dt=0.01$. We have taken the radius of the drop as ${R_0}=200$ when time $t = 0$ and $R^*=50$ when time $t = t_d$. Here $\Gamma=1.0$ and $\epsilon=-0.375$.  

\section{Results and discussions}
The final configuration of the solute depends on its final value of $\psi_{av}$. As mentioned before, if it is within the range of the values of the phase diagram {\cite{grant}} where we get a periodic triangular solid, the resulting structure will be ordered. If it is below the range we get liquid-like disordered state and otherwise we get the stripe phase, which is unphysical for our context. We typically begin the drying process with an initial low solute density or small value of $\psi_{av}$, and this value and the drying rate together determines the final structure.
\begin{figure}[h]
\begin{center}
\includegraphics[width=1.0\textwidth]{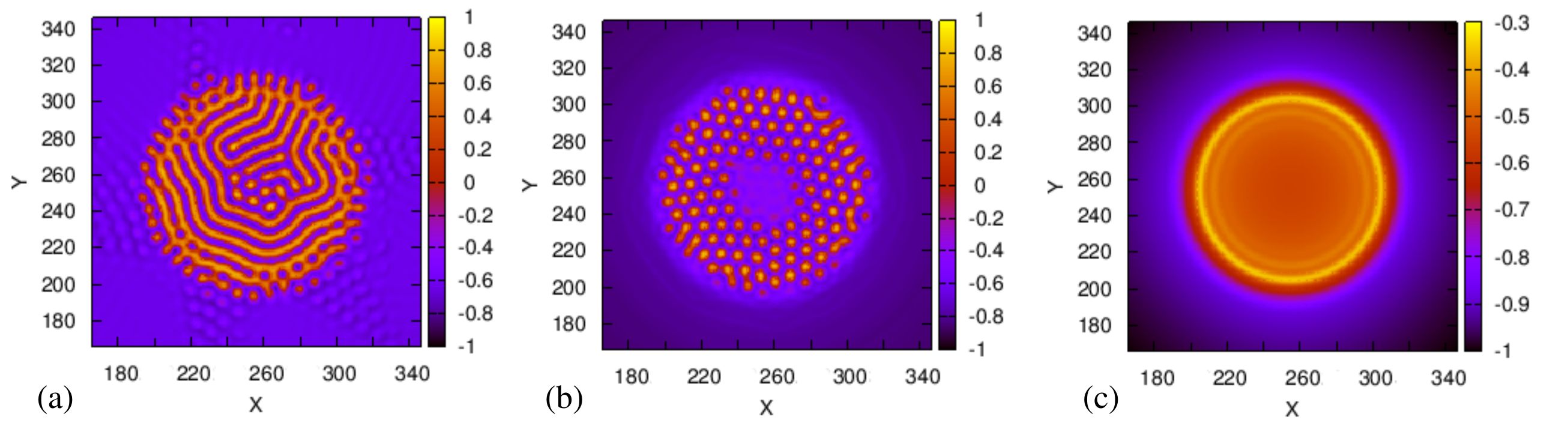}
\caption{ (a),(b) and (c) are the final configuration of $\psi ({\bf r},t)$ for $\psi_{av}^a=-0.7$, $\psi_{av}^b=-1.0$ and $\psi_{av}^c=-1.2$ respectively. All the above three figures \cite{ganai} are for $t_d = 1568.15, h=0.006$ and $\Gamma=1.0$ and they are showing three phases,  striped phase, ordered (triangular) and disordered liquid-like phases respectively. (Using Method I) }
\label{Fig2}
\end{center} 
\end{figure}
\begin{figure}[h]
\begin{center}
\includegraphics[width=0.750\textwidth]{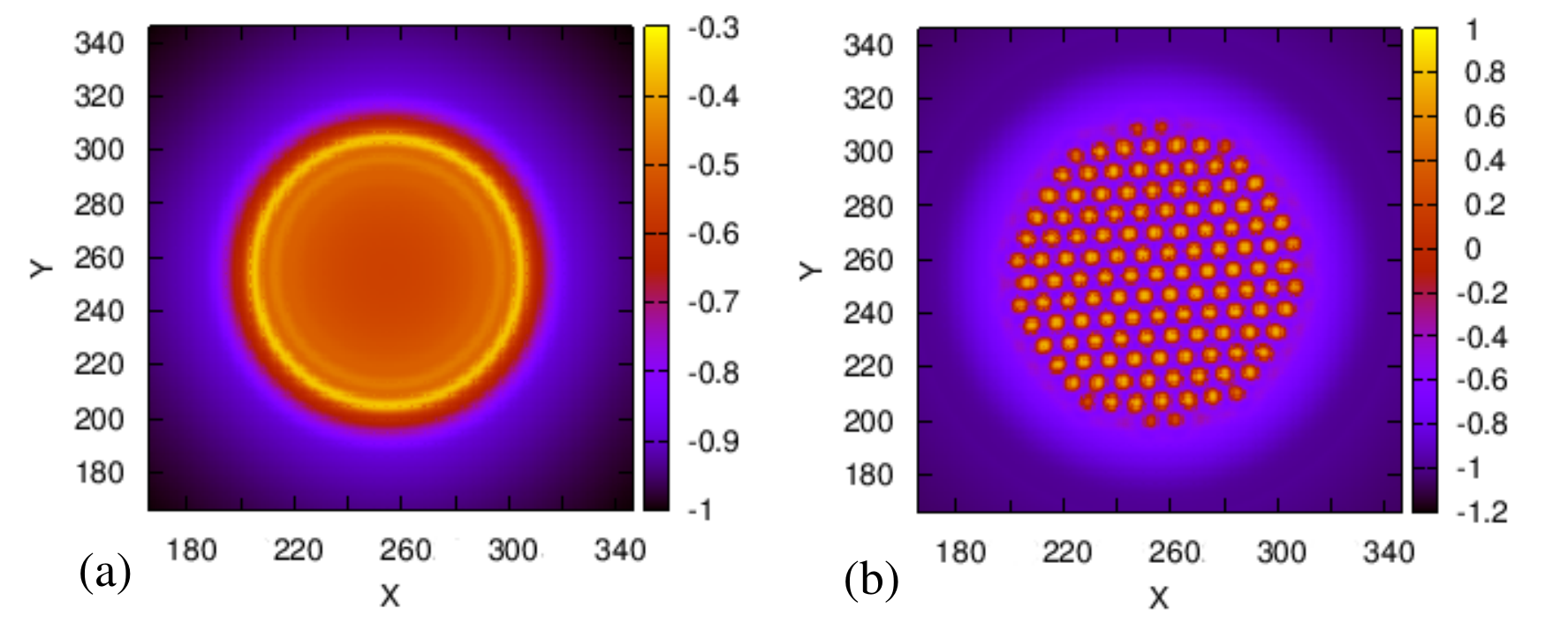}
\caption{ (a) final configuration of $\psi ({\bf r},t)$ for $\psi_{av}=-1.2, t_d = 1568.15, h=0.006$ and $\Gamma=1.0$ (fast drying) and (b) final configuration of $\psi ({\bf r},t)$ for $\psi_{av}=-1.2, t_d = 6620.32, h=0.002$ and $\Gamma=1.0$ (slow drying). For same $\psi_{av}$ but for different drying rate (a) shows liquid-like disordered phase and (b) shows ordered (tringular) phase \cite{ganai}. (Using Method I)}
\label{Fig3}
\end{center} 
\end{figure}

We show typical final configurations of $\psi$ for various initial density and drying rate. Both the methods described earlier, produce similar results. While Fig.\ref{Fig2}(a) shows a striped pattern in the final configuration of the solute, Fig.\ref{Fig2}(b) and Fig.\ref{Fig2}(c) show ordered and liquid-like final configurations respectively. Here the initial values of $\psi_{av}$ are taken according to the order: $\psi_{av}^a > \psi_{av}^b > \psi_{av}^c$ where the superscripts denote respective figures. The total drying time $(t_d)$ in all these cases are same. This implies that for a fixed drying rate and for small initial densities, they are in liquid-like disordered phase and for large initial densities they are in stripe phase. We get crystalline phase in the intermediate initial densities. Fig.\ref{Fig3} (a) shows a typical disordered liquid-like configuration for fast drying and Fig.\ref{Fig3} (b) shows a typical ordered (triangular phase) configuration for slow drying where both have $\psi_{av} = -1.2$ initialy. So, for same initial solute density we get crystalline order when the drying process is slow and liquid-like disordered structure when the drying process is fast. The above results are obtained using method I for solving the dynamical equations.

\begin{figure}[h]
\begin{center}
\includegraphics[width=0.750\textwidth]{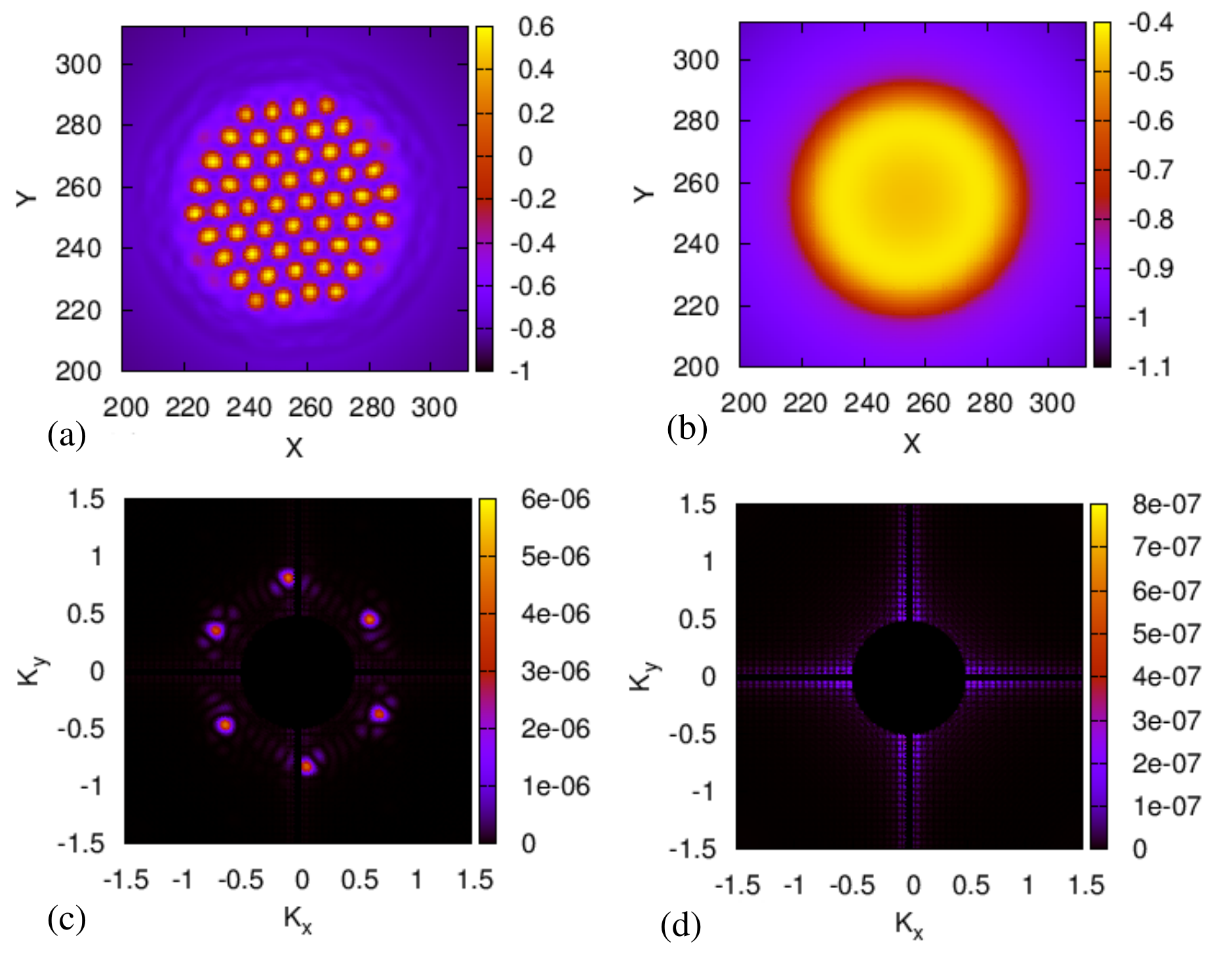}
\caption{ (a) and (b) are the final configuration of $\psi ({\bf r},t)$ for $\psi_{av}^a=-1.1$ and $\psi_{av}^b=-1.3$ respectively. All the above figures are for same $t_d, h, \Gamma$ which are $t_d = 1505.8, h=0.006$ and $\Gamma=1.0$. For same drying rate but for different $\psi_{av}$ (a) shows the ordered phase and (b) shows the disordered phase. (c) and (d) are the structure factor of (a) and (b) respectively. (Using Method II)}
\label{fig4}
\end{center} 
\end{figure}

When method II is used, the results are quite similar validating our ansatz for $\phi$. In Fig.\ref{fig4} we show typical results showing ordered (Fig.\ref{fig4}(a)) and disordered structures (Fig.\ref{fig4}(b)) produced by varying $\psi_{av}$. We have also shown the structure factors corresponding to these configurations in Fig.\ref{fig4}(c) and (d) respectively. We have calculated the structure factors ($S({\bf k})$) using, 
\begin{equation}
S({\bf k}) = {\Big \lbrace \int \,\, \psi({\bf r}) \cos ({\bf k}{\bf .} {\bf r})\,\,d{\bf r} \Big \rbrace }^2 + {\Big \lbrace \int \,\, \psi({\bf r}) \sin ({\bf k}{\bf .} {\bf r})\,\,d{\bf r} \Big \rbrace }^2 
\end{equation}

\begin{figure}[h]
\begin{center}
\includegraphics[width=0.750\textwidth]{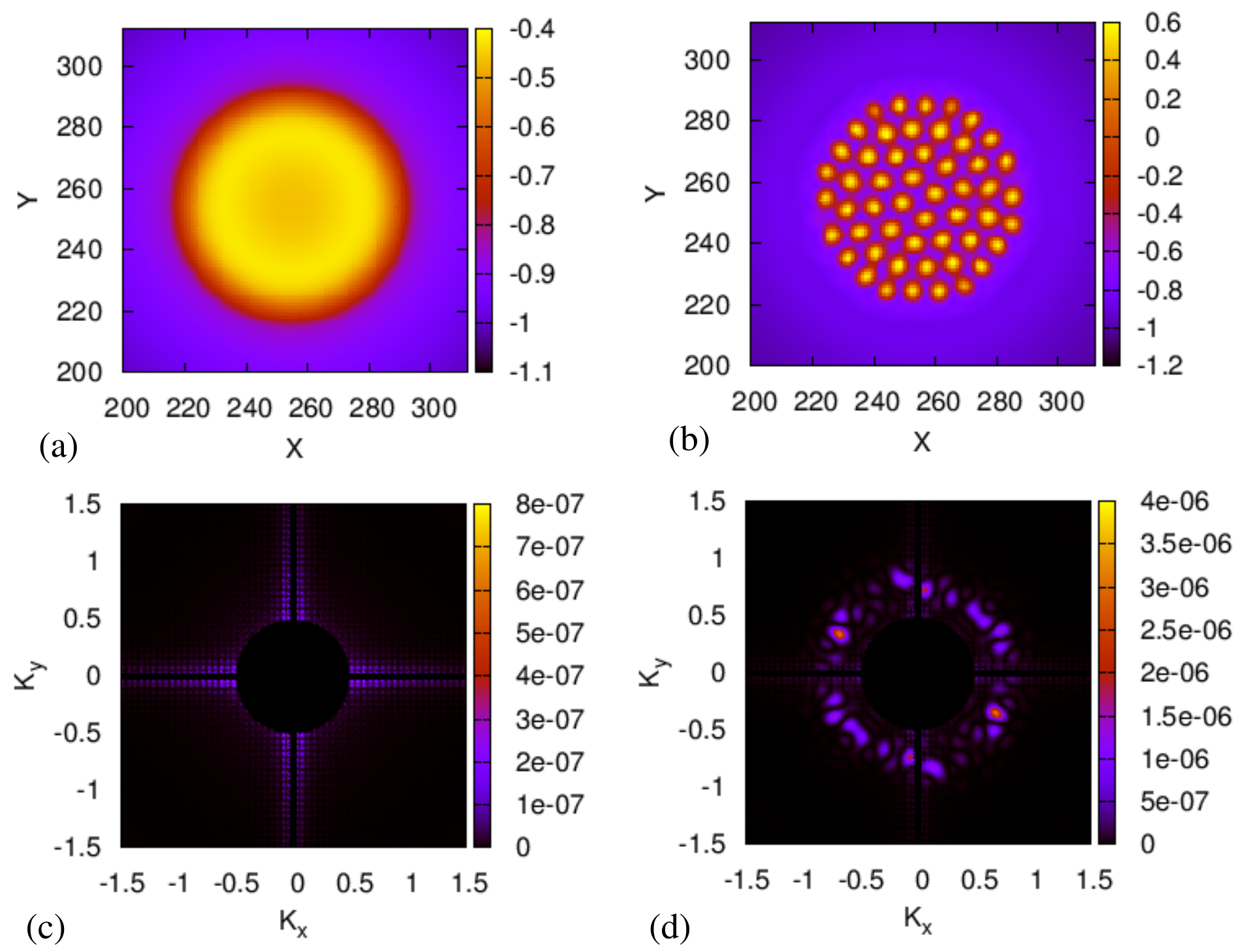}
\caption{ (a) final configuration of $\psi ({\bf r},t)$ for $\psi_{av}=-1.3, t_d = 1505.8, h=0.006$ and $\Gamma=1.0$ (fast drying) and (b) final configuration of $\psi ({\bf r},t)$ for $\psi_{av}=-1.3, t_d = 1810.26, h=0.002$ and $\Gamma=3.0$ (slow drying). For different drying rate but for same $\psi_{av}$ liquid-like disordered configuration is found in (a) and ordered (triangular) configuration is found in (b). (c) and (d) are the structure factor of (a) and (b) respectively. (Using Method II)} 
\label{fig5}
\end{center} 
\end{figure}

In Fig.\ref{fig5}, similar results showing ordered and disordered final configurations (and the corresponding structure factors) obtained by varying the drying rate keeping $\psi_{av}$ fixed are shown. Note that both $\Gamma$ and $h$ has been varied here to change drying time. We have seen that the results depend only on the drying rate and not on $h$ or $\Gamma$ independently.  




\begin{figure}[h]
\begin{center}
\includegraphics[width=0.50\textwidth]{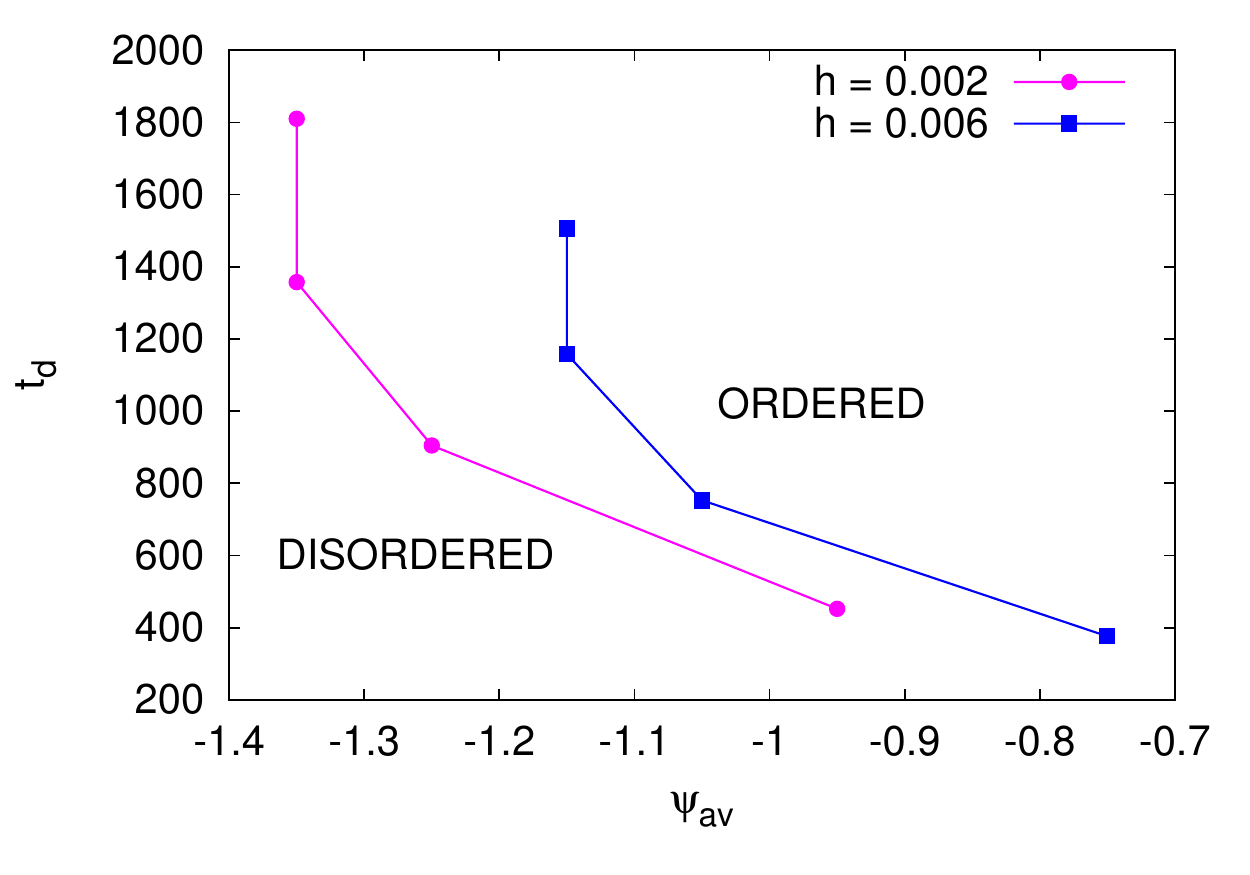}
\caption{ Phase boundary of order-disorder transition for two different $h$ i.e., drying rate, in the parameter space of total drying time of the solvent and the average density of the solute.}
\label{fig6}
\end{center} 
\end{figure}

From the previous discussions, we understand that drying rate of the solvent and initial density of the solute (both of which are externally controllable parameters of the problem) influences the final structure of the solute through a dynamical order-disorder transition. Here we concentrate on the order-disorder phase boundary of the problem. As we are interested only in the two dimensional order-disorder transition, we do not allow the dynamics to take the system into the stripe phase, which is unphysical to the present context. To ensure this, we take recourse to the following artifice. We make $\mu$ not only dependent on $\phi$, but also on $\psi$ mimicking the slowing down of dynamics in the solid phase. We set a cutoff value for $\psi$ above which $\mu=0.0$ and below which $\mu$ satisfies equation(\ref{mueq}). This ensures that the value of $\psi_{av}$ is small enough so that at the end of drying, we have only crystalline or liquid-like disordered state, and not the stripe phase.  When the total mass of the solvent i.e., integrated value of $\phi$ over space, is less than a preset cut-off value representing the detectability limit for solvent, we stop the drying simulation and record the final configuration of $\psi({\bf r},t)$. The time required to arrive at the final stage from the initial configuration, is the total drying time ($t_d$). Averaging over many independent initial configurations, finally we obtain the order-disorder phase boundary in the parameter space of initial solute density and total drying time, which is shown in Fig.\ref{fig6}. To construct the phase boundary, we have used several drying rates obtained by considering two different values of chemical potential: $h=0.002, 0.006$ and four different values of $\Gamma$ and for each of these values, we have also considered several initial $\psi_{av}$ values of the solute. The shape of the phase boundary, shown in Fig.\ref{fig6}, shows that as we increase the solute density, less time is required for drying in order to get the final ordered state.   

\section{Generalization to a binary mixture} In this section we generalize the PFC model of drying induced ordering to a symmetric binary mixture of two species of colloidal particles which interact with one another. We aim to see under what conditions an ordered crystalline arrangement of the colloids is obtained. Needless to say, an ordered arrangement of two species of colloids has many possible technological applications {\cite{soukoulis}}. 

A binary mixture can be modeled by a free energy with minimal coupling of the form \cite{haataja}
\begin{eqnarray}
{\cal F}_3 &=& \int d^2{\bf r} \,\, \Big \lbrace \frac {{\psi}_1}{2} [\epsilon + (1 + \nabla^2)^2] {\psi}_1 + \frac {{\psi}^4_1}{4} + \frac {{\psi}_2}{2} [\epsilon + (1 + \nabla^2)^2] {\psi}_2  \nonumber \\
&& +\: \frac {{\psi}^{4}_{2}}{4} + \alpha {\psi}_1 {\psi}_2 + \phi ({\psi}_1 + {\psi}_2) \Big \rbrace
\label{bfenergy}
\end{eqnarray}

Here the parameter $\alpha$ couples the two fields $\psi_1$ and $\psi_2$ representing the two species of colloidal particles. Note that in Eq.\ref{bfenergy} the two fields are treated on the same footing and the free energy is unchanged under a switch between the two $\psi_i$s. For $\alpha > 0$, the system prefers $\psi_1$ and $\psi_2$ to peak at different locations which leads to a decrease of the overall free energy. Positive $\alpha$ therefore favors ordering. On the other hand if $\alpha$ is negative, the two species of particles tends to occupy, statistically, the same positions within the lattice and therefore denotes a disordered crystal where the two species of particles randomly occupy lattice sites with equal probability - a randomly substituted binary alloy. 

In order to determine the ground states of the above free energy for the (infinite size) binary system we consider the following ansatz  for $\psi_1$ and $\psi_2$. 

\begin{enumerate}
\item {\it (A) Randomly substituted triangular lattice}: This is the structure closest to the single component case where the two species of particles occupy sites of a triangular lattice randomly with equal probability. In this case $\psi_1$ and $\psi_2$ have identical forms given by:
\begin{equation}
\psi_1 = \psi_2 = A_t[\cos(q_tx)\cos(q_ty/\sqrt{3}) - \cos(2q_ty/\sqrt{3})/2] + \psi_{av}
\label{triangular}
\end{equation}

\item {\it (B) Two sublattice ordered square lattice}: This structure consists of a two dimensional square lattice which is bipartite and hence can be decomposed into two sublattices. One sublattice is occupied by $\psi_1$ whereas the other is occupied by $\psi_2$. The corresponding forms for $\psi_1$ and $\psi_2$ are:
\begin{eqnarray}
\psi_1 &=& A_s[\cos(q_sx) + \cos(q_sy)] + \psi_{av} \nonumber \\
\psi_2 &=& A_s[\cos(q_s(x + a/2)) + \cos(q_s(y + a/2))] + \psi_{av}
\label{square}
\end{eqnarray}

\item {\it (C) Triangular lattice with three sublattice order}: It is difficult to make an ordered arrangement of two species of particles on a triangular lattice which is {\it tripartite} due to geometric frustration. In this case, the problem is solved by keeping one of the sublattices vacant and $\psi_i$ occupying the other two leading to a relatively open honeycomb like structure. The functional forms of $\psi_1$ and $\psi_2$ are given by: 
\begin{eqnarray}
\psi_1 &=& A_{st}[\cos(q_{st}x)\cos(q_{st}y/\sqrt{3}) - \cos(2q_{st}y/\sqrt{3})/2] + \psi_{av} \nonumber  \\
\psi_2 &=& A_{st}[\cos(q_{st}(x + a/2))\cos(q_{st}(y + a/2\sqrt{3})/\sqrt{3}) \nonumber \\
&& -\: \cos(2q_{st}(y + a/2\sqrt{3})/\sqrt{3})/2] + \psi_{av} 
\label{striangular}
\end{eqnarray}

In the above three equations $A_t, A_s, A_{st}$ are unknown constants and $q_t = q_s = q_{st} = 2\pi/a$ where $a$ is the lattice constant.

\end{enumerate}

\begin{figure}[h]
\begin{center}
\includegraphics[width=0.75\textwidth]{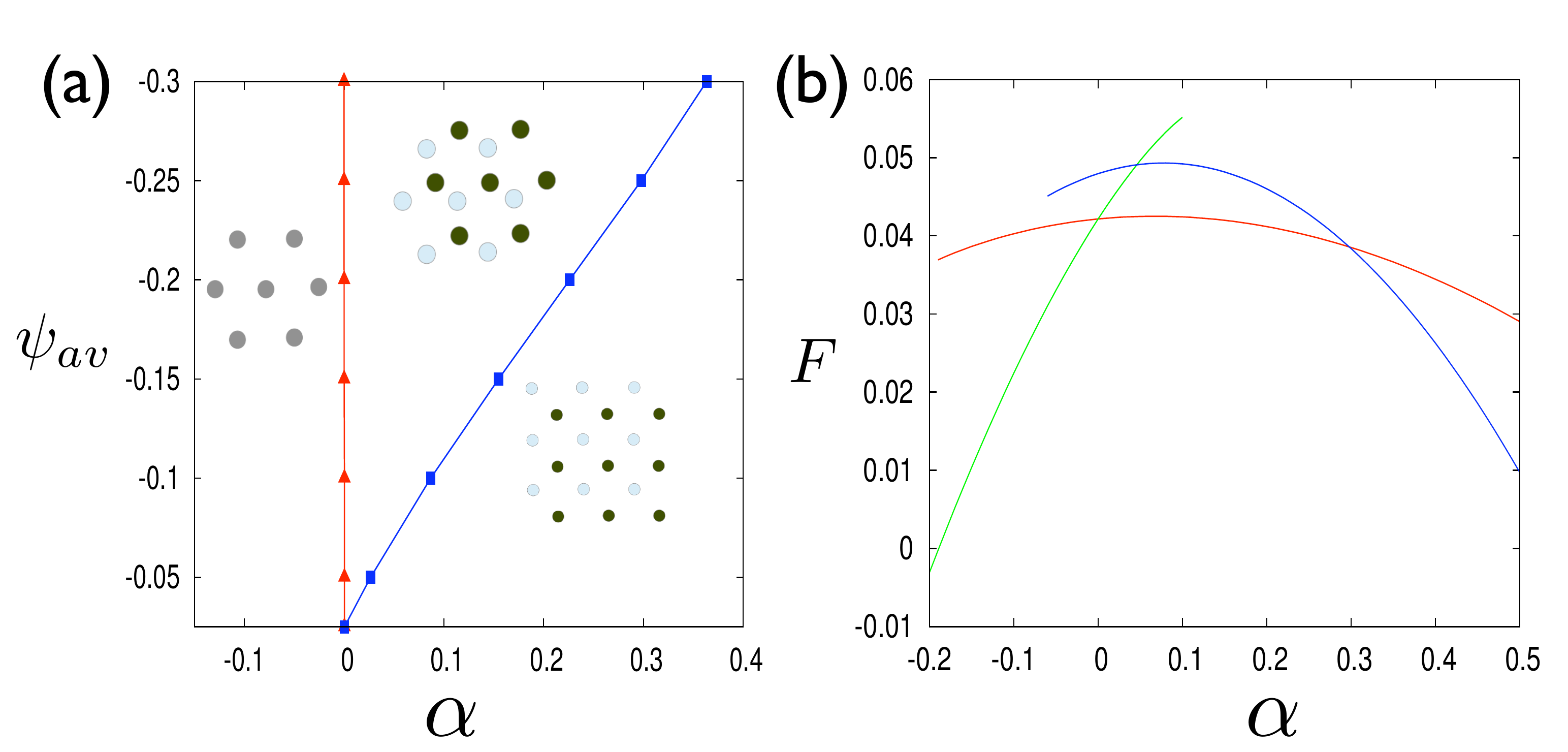}
\caption{ (a) Phase diagram in the $\psi_{av} - \alpha$ plane of the three different lattice structures. The red line with filled triangle is the phase boundary between randomly substituted triangular lattice and triangular lattice with one sublattice vacant and the blue line with filled square is that between triangular lattice with one sublattice vacant and two sublattice ordered square lattice. (b) plot of free energies of the three different structures per unit area as a function of $\alpha$ for $\psi_{av} = -0.25$. Green, red and blue colour lines represent randomly substituted triangular lattice,triangular lattice with one sublattice vacant and two sublattice ordered square lattice respectively.}
\label{fig7}
\end{center} 
\end{figure}

\begin{figure}[h]
\begin{center}
\includegraphics[width=1.0\textwidth]{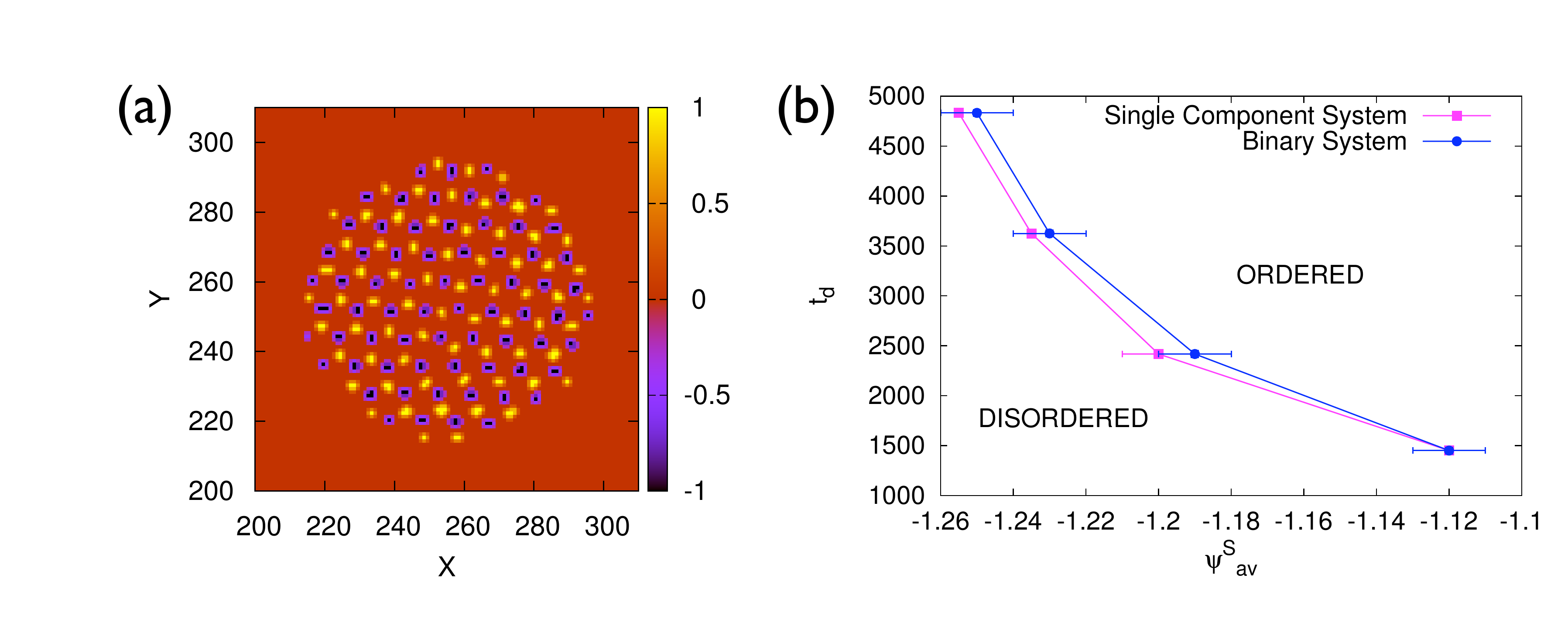}
\caption{ (a) Plot of the maxima of $\psi_1$ and $\psi_2$ showing the three sub-lattice order. Colors yellow and blue corresponds to the maxima of $\psi_1$ and $\psi_2$ respectively. Here average value of $\psi_1$ and $\psi_2 = -0.6$, $\Gamma = 1.0$, $h = 0.006$ and total drying time $t_d = 1450.0 $ (b) Comparison between phase boundaries of order-disorder transition of single component system and binary system for $h = 0.006$ in the parameter space of total drying time ($t_d$) of the solvent and average density of the solute. For binary system, $\psi^{S}_{av}$ is the intial average density of one component of the binary system (i.e. $\psi_1$ or $\psi_2$).}
\label{fig8}
\end{center} 
\end{figure}

%

Substituting Eq.\ref{triangular} into Eq.\ref{bfenergy} and after integrating over unit cell and minimising with respect to $A_t$ and $q_t$, we get free energy per unit area of the randomly substituted triangular lattice as 
\begin{eqnarray}
\frac {F_t}{S} &=& \frac{45}{256}{A_t}^4 - \frac{3}{8}\psi_{av}{A_t}^3 + \frac{{A_t}^2}{8}[9{\psi_{av}}^2 + 3(\epsilon + \alpha)] \nonumber \\
&& +\: (\epsilon + \alpha + 1){\psi_{av}}^2 + \frac{{\psi_{av}}^4}{2}
\label{tfe}
\end{eqnarray} 
where $A_t = \frac{4}{5} \Big [\psi_{av} - \frac{1}{3}\sqrt{-36{\psi_{av}}^2 - 15(\epsilon + \alpha)} \Big ]$, $q_t = \frac{\sqrt{3}}{2}$ and $S$ is the area of the unit cell. Similarly, substituting Eq.\ref{square} into Eq.\ref{bfenergy}, free energy per unit area of the two sublattice ordered square lattice can be written as
\begin{equation}
\frac {F_s}{S} = \frac{9}{8}{A_s}^4 + {A_s}^2[3{\psi_{av}}^2 + \epsilon - \alpha] + (\epsilon + \alpha + 1){\psi_{av}}^2 + \frac{{\psi_{av}}^4}{2}
\label{sfe}
\end{equation} 
where $A_s =  \frac{2}{3}\sqrt{-3{\psi_{av}}^2 - \epsilon + \alpha} $, $q_s = 1$ and finally, substituting Eq.\ref{striangular} into Eq.\ref{bfenergy}, free energy per unit area of the triangular lattice with one sublattice vacant becomes
\begin{eqnarray}
\frac {F_{st}}{S} &=& \frac{45}{256}{A_{st}}^4 - \frac{3}{8}\psi_{av}{A_{st}}^3 + \frac{{A_{st}}^2}{8}[9{\psi_{av}}^2 + 3(\epsilon - \alpha/2)] \nonumber \\
&& +\: (\epsilon + \alpha + 1){\psi_{av}}^2 + \frac{{\psi_{av}}^4}{2}
\label{stfe}
\end{eqnarray} 
where $A_{st} = \frac{4}{5} \Big [\psi_{av} - \frac{1}{3}\sqrt{-36{\psi_{av}}^2 - 15(\epsilon - \alpha/2)} \Big ]$, $q_t = \frac{\sqrt{3}}{2}$. 
In Fig.\ref{fig7}(b) we have plotted the free energies of the three structures per unit area as a function of the parameter $\alpha$ for $\psi_{av} = -0.25$. It is reflected from the Fig.\ref{fig7}(b) that for $\alpha < 0$, randomly substituted triangular lattice is more probable than the other structures because of its lower value of free energy but for $0 < \alpha < 0.3$ triangular lattice with one sublattice vacant has lower free energy value and for $\alpha > 0.3$ two sublattice ordered square lattice has the smaller free energy value. Fig.\ref{fig7}(a) shows the phase diagram of the three different lattice structures in $\psi_{av} - \alpha$ plane. It is obtained by calculating free energies of different lattices per unit area for different values of $\psi_{av}$ and $\alpha$ i.e., putting different values of $\psi_{av}$ and $\alpha$ in Eq.\ref{tfe}, Eq.\ref{sfe} and Eq.\ref{stfe}. A global minimization of the free energy among the three chosen structures then yields the phase diagram. Fig.\ref{fig7}(a) also informs that randomly substituted triangular lattice is more probable in the region where $\alpha < 0$ and for slightly higher positive value of $\alpha$ two sublattice ordered square lattice has the greater chance to appear and in between these ranges of $\alpha$, triangular lattice with one sublattice vacant is more probable.

Having determined the phase diagram for the binary system, we are now in a position to address the problem of drying induced ordering of the binary mixture. The dynamical equations for $\psi_i$ may be written down in conformity with the single component case as, 
\begin{equation}
\frac {\partial \psi_1}{\partial t} =  \overrightarrow{\nabla} {\bf .} (\mu \overrightarrow{\nabla} \frac {\delta {\cal F}_3}{\delta \psi_1})
\label{em1}
\end{equation}
\begin{equation}
\frac {\partial \psi_2}{\partial t} =  \overrightarrow{\nabla} {\bf .} (\mu \overrightarrow{\nabla} \frac {\delta {\cal F}_3}{\delta \psi_2})
\label{em2}
\end{equation}

In our case, $\alpha = 0.1$ is taken. We also considered here a binary system having equal average density of each component. And we found that when each component formed an ordered configuration then the density peak of each component does not coincide each other in space which is clearly seen in the Fig.\ref{fig8}(a). For this value of $\alpha$, the final configuration is that of an ordered triangular lattice with one sublattice vacant (C). We have also tried with higher value of $\alpha$ (positive), where a two sublattice ordered square lattice is expected. However, in this regime we always obtain a disordered configuration and no equilibrium square lattice. We suspect that this is because the circular boundary in our system is more compatible with a triangular rather than a square lattice, a situation compounded by the (necessarily) small system size. 

The dynamical phase boundary of the order-disorder transition of binary system in the parameter space of total drying time of the solvent and total average density of the solute is shown in Fig.\ref{fig8}(b) which also indicates, like single component system, that as we increase the average density of solute, less time is required for drying to get the ordered state finally. We know that total drying time can be varied by changing the value of chemical potential $h$ and mobility $\Gamma$ but here total drying time of solvent is varied by only changing mobility $\Gamma$, keeping chemical potential $h$ fixed ($h = 0.006$). We have also Plotted the phase boundary of the order-disorder transition of a single component system in the same figure (Fig.\ref{fig8}(b)) showing that, qualitatively, the two systems are similar to each other.  

\section{Discussion and future directions}

In the real system\cite{DD}, the dynamical phase boundary is not monotonic as in our studies (Figs.\ref{fig6} and \ref{fig8}(b)), but re-entrant signifying that for higher densities one obtains, again, a disordered final state. The reason for this, as explained in Ref.\cite{DD} is that there are two timescales in the problem. The first timescale corresponds to the time required to compactify the system and this decreases with increasing density. The second timescale corresponds to the nucleation time for the system to find the low free energy ordered state in configuration space. The second timescale increases with density due to an impending glass transition. In the simple versions of the PFC model that we have used, there is no glassy solution present and hence we cannot model this second phenomenon within our approach in any simple manner. There has however been some work in trying to generalize the PFC model to include glassy states \cite{{berry1},{berry2}} and slow dynamics. We hope to extend our work in this direction in the near future. 

Apart from spherical particles (whether monodisperse or bidisperse) one may try to study the drying of non-spherical particles by suitably modifying the PFC model. In this case one may need to invoke additional order parameters such as the local director. Similarly, in order to incorporate hydrodynamic effects, additional fields corresponding to the momentum densities of the solute and the solvent need to be included. Work along these lines are in progress.   


{\bf Acknowledgements.\,}    
Funding from the Indo-EU project MONAMI is gratefully acknowledged.

\end{document}